\documentclass[aps,preprint,amsfonts,amsmath,endfloats]{revtex4}

\usepackage{graphicx}  % Graphics capability
\usepackage{color}

\begin{document}
\author{Martin J.~Greenall}
\affiliation{Institut Charles
  Sadron, 23, rue du Loess, 67034 Strasbourg, France}
\email{mjgreenall@physics.org}
\author{Gerhard Gompper}
\affiliation{Theoretical Soft Matter and Biophysics, Institute for Complex Systems,
 Forschungszentrum J\"{u}lich, 52425 J\"{u}lich, Germany}
\title{Simple and complex micelles in amphiphilic mixtures: a
  coarse-grained mean-field study}
\begin{abstract}
Binary mixtures of amphiphiles in solution can self-assemble into a wide range
of structures when the two species
individually form aggregates of different curvatures. In this paper,
we focus on small, spherically-symmetric aggregates in a solution of
sphere-forming amphiphile mixed with a smaller amount of
lamella-forming amphiphile. Using a coarse-grained mean-field model
(self-consistent field theory, or SCFT), we scan the parameter space
of this system and find a range of morphologies as the interaction
strength, architecture and mixing ratio of the amphiphiles are varied. When the two
species are quite similar in architecture, or when only a small amount
of lamella-former is added, we find simple spherical
micelles with cores formed from a mixture of the hydrophobic blocks of
the two amphiphiles. For more
strongly mismatched amphiphiles and higher lamella-former
concentrations, we instead find small vesicles and more complex micelles. In
these latter structures, the lamella-forming species is encapsulated by the
sphere-forming one. For certain interaction strengths and
lamella-former architectures, the amount of lamella-forming copolymer
encapsulated may be large, and the implications of this for the
solubilization of hydrophobic chemicals are considered. The mechanisms
behind the formation of the above structures are discussed, with a
particular emphasis on the sorting of amphiphiles according to their
preferred curvature.
\end{abstract}
\maketitle
\section{Introduction}

Amphiphilic molecules such as block copolymers and lipids can self-assemble into many different structures when
dissolved in solution
\cite{jain_bates,battaglia_ryan}. This phenomenon
has recently attracted a great deal of attention \cite{smart,howse},
driven both by the potential applications of self-assembled
amphiphile aggregates in the encapsulation and delivery of active
chemicals such as drugs and
genetic material \cite{kim,lomas} and the insights gained into
biological systems \cite{zidovska}.

For solutions of a single type of simple amphiphile, such as a diblock
copolymer or simple lipid, it is fairly
straightforward to gain a basic understanding of which aggregate will
form in a given system \cite{kinning_winey_thomas,israelachvili}. Although a
variety of factors, such as the concentration
\cite{kinning_winey_thomas,adams} and size \cite{kaya} of the
amphiphilic molecules, may play a role, the shape of the aggregates is
most easily controlled via the architecture
of the amphiphile; that is, the relative sizes of its hydrophilic and
hydrophobic blocks. If the hydrophilic component is large (or appears
large due to its interaction with the solvent), then spherical
micelles are seen. However, if the hydrophobic block is large,
lamellar structures such as vesicles form. For intermediate
architectures, cylindrical micelles are observed, either as isolated,
worm-like structures \cite{schuetz}, or branched networks \cite{dan_safran}.

The experimental phenomenology is much richer in binary mixtures of
amphiphiles \cite{kaler}, especially those
that individually self-assemble into different aggregates
\cite{jain_bates_macro,safran_pincus_andelman,safran_et_al}. Novel structures are observed, such as 
undulating cylinders and branched, octopus-like aggregates
\cite{jain_bates_macro}. Binary mixtures have been investigated in a
wide variety of amphiphile systems. A great deal of work has been
carried out on lipid-detergent systems \cite{vinson,oberdisse}, and
over the past few years lipid \cite{sorre,zidovska,gg} and block copolymer
\cite{jain_bates_macro,schuetz} mixtures have been widely studied. Lipid mixtures are of interest due to their presence
in cells and role in biological transport processes \cite{akiyoshi}.  In the case of
block copolymers, on the other hand, the motivation for the use of two amphiphiles is
that it greatly increases the number
of design parameters and gives finer control over the
self-assembly. The architectures and concentrations of both species
may now be varied, as may the stage in the self-assembly process at
which they are blended
\cite{schuetz}. A number of properties of the aggregates may be
controlled, such as their shape \cite{schuetz}, stability \cite{lee}, and
ability to solubilize hydrophobic compounds \cite{oh}. An interesting
and recent example is the
addition of lamella-forming copolymers with a
short hydrophilic block to a solution of longer sphere-formers to
increase the solubilization capacity of the resulting micelles while
maintaining their compact and stable spherical shape
\cite{lee,oh}. In the current paper, we study a basic example of such
a system: a solution of sphere-forming diblock copolymers to which an admixture of
diblocks with a much shorter hydrophilic block is added. To study the problem in as simple a
form as possible, we consider two copolymer species that are
formed of the same species A (hydrophilic) and B (hydrophobic), and
have the same length hydrophobic blocks. We focus on the case where the
sphere-formers remain in the majority, and, using coarse-grained
mean-field theory, investigate how the small
spherical aggregates formed are modified by the presence of the
shorter copolymers. We perform a broad scan of the system's parameter space, and
study how the core composition and radius of the micelles are affected
by the interactions, concentrations and architectures of the two
polymers.

The paper is organized as follows. In the following section, we
introduce the coarse-grained mean-field theory (self-consistent field
theory) that will be used. We then present and discuss our theoretical results,
and give our conclusions in the final section.

\section{Self-consistent field theory}\label{scft}

Self-consistent field theory (SCFT) \cite{edwards} is a coarse-grained
mean-field model that has been used
successfully to model equilibrium
\cite{maniadis,drolet_fredrickson,matsen_book} and metastable
\cite{duque,katsov1} structures in polymers blends and melts. SCFT has
several features that make it particularly suitable for the study of
the current problem of small binary aggregates. First, its general
advantages are that it is less computationally intensive than simulation
techniques such as Monte Carlo, yet, for sufficiently long amphiphiles
\cite{cavallo}, provides comparably accurate
predictions of micelle size and shape
\cite{cavallo,wijmans_linse,leermakers_scheutjens-shape}. Secondly, as a
relatively simple, coarse-grained theory, it will allow us to model the
broad phenomenology of the system clearly and show
how general the phenomena observed are likely to be. Furthermore,
SCFT has a specific feature that is
especially useful in the current problem: it makes no initial assumption about the segregation of two
copolymers of different architecture within the micelle, provided the
two amphiphile species are formed from the same types
of monomer. This will enable us to demonstrate that effects such as
the encapsulation of one polymer species by another within the micelle
arise spontaneously and do not require further assumptions to be made.

To make our discussion more concrete, we now outline the mathematical
structure and main assumptions of
SCFT as applied to the current system. The theory considers an ensemble
of many polymers. These are modeled as random walks
in space, which means that fine details of their molecular structure
are not taken into account \cite{schmid_scf_rev}. The inter-molecular interactions are
modeled by assuming that the system is incompressible and introducing a
contact potential between the molecules \cite{matsen_book}, the strength of which
is fixed by the Flory $\chi$ parameter \cite{jones_book}.

The first step in finding an approximation method (SCFT) for this
problem is to think of each polymer molecule as being acted on by a field
produced by all other molecules in the system \cite{matsen_book}.
Viewing the problem in this way involves no approximations in itself, but
has several advantages when
computing numerical solutions. First, it transforms the $N$-body
problem of modeling a system of $N$
polymers into $N$ $1$-body problems \cite{matsen_book}. The advantage
of this arises from the fact that,
since we are
interested in computing average properties, we consider the
partition sum over all possible system configurations. This means that all molecules
of a given species may be treated as equivalent and we in fact only have to
solve one $1$-body problem for each type of polymer in the
system. Second, this approach allows us to replace the discrete sum
over the system configurations by a continuous integral over smooth functions,
which is easier to deal with numerically \cite{katsov1}. Finally, the computational
burden of the problem may be sharply reduced by finding approximate forms of
the field variables using a saddle-point (mean-field) approximation
\cite{schmid_scf_rev}, which corresponds to neglecting fluctuations in the system.

SCFT can be used to study a wide variety of polymer systems, including
simple homopolymers \cite{werner}, more complex copolymers
\cite{mueller,wang} and mixtures of these \cite{denesyuk}. We now outline
the application of SCFT to our current system of two
amphiphiles in a solvent, which we model by a simple
mixture of two types of AB block copolymer with A homopolymer solvent. We take the lamella-forming species of
copolymer to have a mean-squared
end-to-end distance of $a^2N$, where $a$ is the monomer length and $N$
is the degree of polymerization \cite{matsen_book}. This polymer
contains $N_\text{B}$ hydrophobic B-monomers and $N_\text{A}=N-N_\text{B}$
hydrophilic A-monomers. To consider the effect of different
lamella-formers on the self-assembly, we vary the number of A monomers
while keeping $N_\text{B}$ fixed.

We note that our use of the term lamella-former to refer to
amphiphiles with short hydrophilic blocks is not precise, as these molecules might
precipitate rather than self-assemble in solution \cite{mayes} if not mixed with an amphiphile with a larger hydrophilic
block such as a sphere-former. Furthermore, in aggregates formed from a
mixture of sphere-formers and amphiphiles with a short hydrophilic
block, the presence of these latter molecules
might lead to the formation of regions of negative curvature rather
than the zero-curvature regions that would be favored by a
lamella-former.

All sphere-formers considered contain
$N_\text{B2}\equiv N_\text{B}$ hydrophobic monomers, but their overall length is
necessarily greater and is given by $\alpha N$, where
$\alpha>1$. As above, the number of A monomers $N_\text{A2}$ is varied with
$N_\text{B}$ fixed in order to investigate the effect of sphere-former
architecture on the aggregate properties.
For simplicity, the number $N_\text{S}$ of A
monomers in a homopolymer solvent molecule is also fixed at
$N_\text{B}$. Since we focus on spherical aggregates, we assume
spherical symmetry of the calculation box with reflecting boundary
conditions at the origin and outer limit of the system.

In this paper, we keep the amounts of copolymer and homopolymer fixed; that is, we work in the canonical ensemble. Applying the procedure described above, we find that the SCFT approximation to the free energy of our system has the form
\begin{align}
\lefteqn{\frac{FN}{k_\text{B}T\rho_0V}=\frac{F_\text{h}N}{k_\text{B}T\rho_0V}}\nonumber\\
& 
-(\chi N/V)\int\mathrm{d}\mathbf{r}\,\left[(\phi_\text{A}(\mathbf{r})+\phi_\text{A2}(\mathbf{r}) +\phi_\text{S}(\mathbf{r})-\overline{\phi}_\text{A}-\overline{\phi}_\text{A2}-\overline{\phi}_\text{S})(\phi_\text{B}(\mathbf{r})+\phi_\text{B2}(\mathbf{r})-\overline{\phi}_\text{B}-\overline{\phi}_\text{B2})\right]
\nonumber\\
& 
-(\overline{\phi}_\text{A}+\overline{\phi}_\text{B})\ln (Q_\text{AB}/V)-[(\overline{\phi}_\text{A2}+\overline{\phi}_\text{B2})/\alpha]\ln
(Q_\text{AB2}/V) -\overline{\phi}_\text{S}\ln (Q_\text{S}/V)
\label{FE}
\end{align}
where the $\overline{\phi}_i$ are the mean volume fractions of the
various components. The $\phi_i(\mathbf{r})$
are the local volume fractions, with $i=A$ or $A2$ for the hydrophilic
components of species $1$ and $2$, $i=B$ or $B2$ for the hydrophobic
components and $i=S$ for the A homopolymer solvent. The strength of the repulsive interaction
between the A monomers (hydrophilic component and solvent) and B
monomers (hydrophobic component) is determined by the Flory parameter $\chi$. $V$ is the
total volume, $1/\rho_0$ is the volume of a monomer, and $F_\text{h}$
is the SCFT free energy of a homogeneous system of the same
composition. This last quantity contains terms arising from the
entropy of mixing of the different species \cite{jones_book} and a term proportional to $\chi$ describing the interactions of the species in the absence of self-assembly \cite{matsen_book}. The architectures of the individual molecules enter through the
single-chain partition functions $Q_i$. As an example, that for the
homopolymer is given by \cite{matsen_book}
\begin{equation}
Q_\text{S}[w_\text{A}]=\int\mathrm{d}\mathbf{r}\,q_\text{S}(\mathbf{r},s)q_\text{S}^\dagger(\mathbf{r},s)
\label{single_chain_partition}
\end{equation}
where the $q$ and $q^\dagger$ terms are single chain propagators
\cite{matsen_book}. The partition functions of the copolymer chains
are calculated in a similar way. We now recall that the
polymer molecules are modeled as random walks in an external
field that describes their interactions with the other molecules in
the system. This is reflected in the fact that the propagators satisfy
diffusion equations with a field term. Again considering the case of the homopolymer, we
have
\begin{equation}
\frac{\partial}{\partial
s}q_\text{S}(\mathbf{r},s)=\left[\frac{1}{6}a^2N\nabla^2-w_\text{A}(\mathbf{r})\right]q_\text{S}(\mathbf{r},s)
\label{diffusion}
\end{equation}
where $s$ is a curve parameter specifying the position along the
polymer and the initial condition is $q_\text{S}(\mathbf{r},0)=1$. The copolymer
propagators are computed similarly, although in this case the
corresponding diffusion equation is solved with
the field $w_i(\mathbf{r})$ and the prefactor of the $\nabla^2q$
term appropriate to each of the two sections of the copolymer
\cite{fredrickson_book}. In the case of the longer sphere-forming
copolymer, the fields must be multiplied by the ratio $\alpha$ to take
into account the higher
degree of polymerization \cite{matsen_book}.

The derivation of the mean-field free energy $F$ also generates a set of simultaneous equations linking
the values of the fields and densities. The first of
these simply states that all volume fractions must add to $1$ at all
points due to the incompressibility of the system;
however, we also find the following linear relation
\begin{equation}
w_\text{A}(\mathbf{r})-w_\text{B}(\mathbf{r})=2\chi N[\overline{\phi}_\text{A}+\overline{\phi}_\text{A2}+\overline{\phi}_\text{S}-\phi_\text{A}(\mathbf{r}) -\phi_\text{A2}(\mathbf{r})-\phi_\text{S}(\mathbf{r})]
\label{SCFT_equation}
\end{equation}
Furthermore, the homopolymer density is related to the propagators
(see Eqn.\ \ref{diffusion}) according to
\cite{matsen_book}
\begin{equation}
\phi_\text{S}(\mathbf{r})=\frac{V\overline{\phi}_\text{S}}{Q_\text{S}[w_\text{A}]}\int^1_0\mathrm{d}s\,
q_\text{S}(\mathbf{r},s)q_\text{S}^\dagger(\mathbf{r},s)
\label{density}
\end{equation}
The copolymer densities are computed in a similar way, with the
integration limits set to give the required amounts of each
species.

In order to calculate the SCFT density profiles for a given set of
polymer concentrations, Eqn.\ \ref{SCFT_equation} must be solved with the densities calculated as
in Eqn.\ \ref{density} and taking account of the incompressibility of the
system. To begin, we make a initial guess
for the fields $w_i(\mathbf{r})$ with the approximate form
of a micelle and solve the diffusion
equations to calculate the propagators and then the densities
corresponding to these fields (see Eqns \ref{diffusion} and
\ref{density}). New values for the fields are now calculated using
the new $\phi_i(\mathbf{r})$, and the $w_i$ are updated
accordingly \cite{matsen2004}. The procedure is repeated until convergence is achieved.

The diffusion equations are solved using a finite difference method
\cite{num_rec}. To resolve the more complex features of the micelle
density profiles, it is necessary to use a relatively fine
discretization: a spatial step size of $0.028\,aN^{1/2}$ and a step
size for the curve parameter $s$ of $0.0025$. 

Until now, we have only considered isolated spherical micelles. We
must now link the thermodynamics of a single aggregate to those of a larger
system containing many micelles. To do this, we proceed as follows \cite{gbm_macro,gbm_jcp}. Firstly, we
calculate the free-energy
density of a box containing a single spherical aggregate
surrounded by solvent. Since we assume spherical symmetry, the calculation is effectively one-dimensional.
The volume of the simulation box
containing the aggregate is then varied, keeping the volume fraction of copolymer
constant, until the box size with the minimum free-energy density is found.
Provided the system is dilute, so that micelle is surrounded by a large volume of solvent, this
mimics the behavior of a larger system (of fixed
volume and fixed copolymer volume fraction) containing many
micelles. The reason for this is that such a system minimizes its total
free energy by changing the number of micelles and therefore the volume
(`box size') occupied by each. Minimizing the free energy density in this way
locates the micelle that would have the lowest free energy and hence
be most likely to be observed in a sample containing many aggregates.
This approach allows many-micelle systems to be investigated using inexpensive one-dimensional
calculations on single aggregates, and its predictions on micelle
radii and shape transitions often agree well with experiment \cite{gbm_macro,gbm_jcp}. 

\section{Results and discussion}\label{results}
We divide the results section into four subsections. These focus on
the effect on the micelle morphology of, respectively, the lamella-former
concentration, the strength of the interaction between the two
species, the lamella-former architecture and the sphere-former architecture.
\subsection{Effect of lamella-former concentration}
To begin, we investigate the effect of the gradual addition of
lamella-former to a system of sphere-formers, finding the micelle of lowest
free energy using the method of variable subsystem size described
above. To see the effect of blending two species as clearly as
possible, we consider a pair of strongly mismatched copolymers: a
lamella-former with $N_\text{A}=N_\text{B}/4$ and a sphere-former with
$N_\text{A2}=7N_\text{B}$.
The Flory parameter is set to
$\chi N_\text{B}=22.5$, a high value which causes relatively sharp
interfaces to form between the species A and B. We fix the overall volume fraction of
copolymer to $10\%$, to give a reasonable volume of solvent around
the micelle without making the simulation box so large that the
calculations become slow.

\begin{figure}
\includegraphics[width=\linewidth]{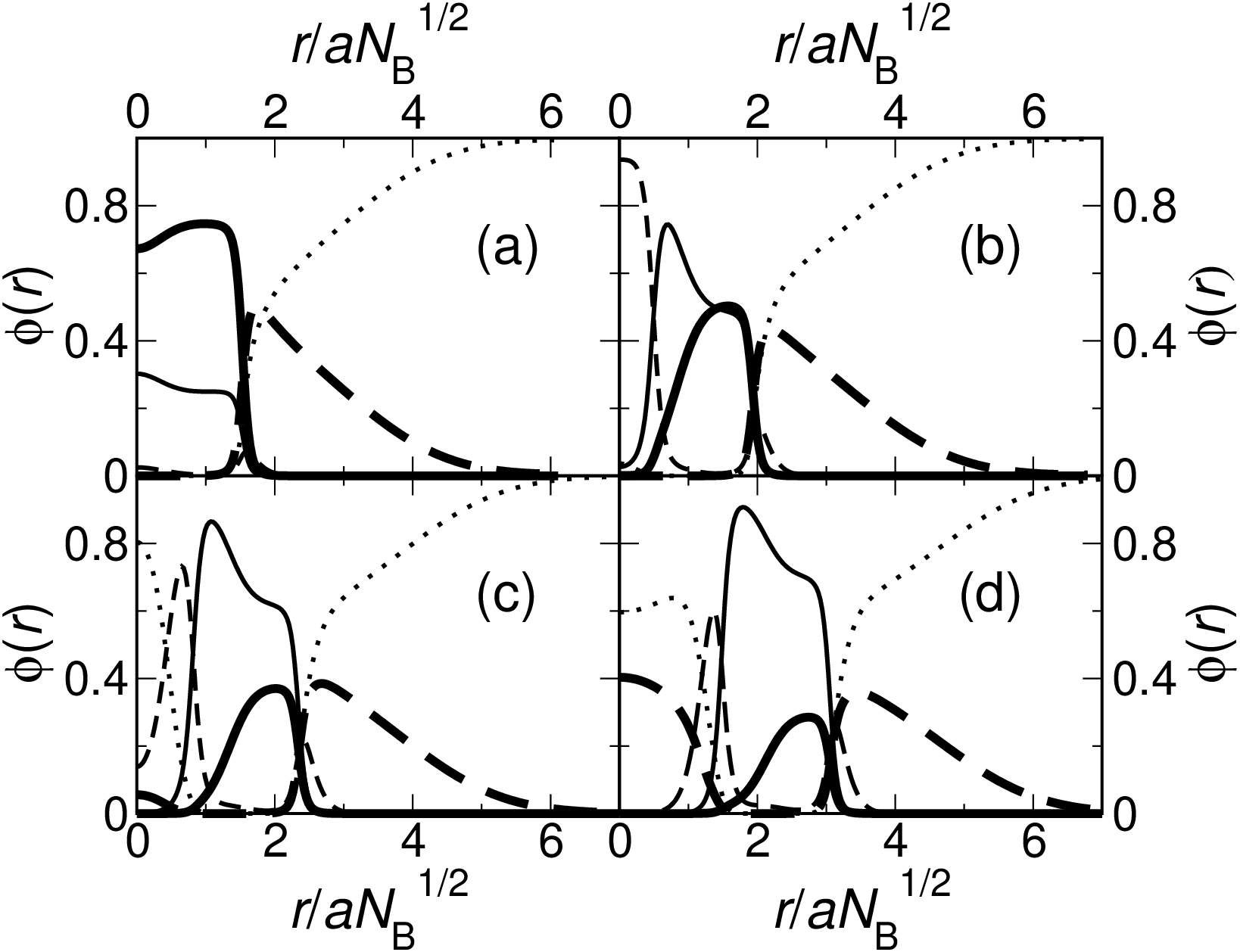}
\caption{\label{conccut45_fig} Cuts through the density profiles of the
  spherically-symmetric aggregates formed in a solution of
  lamella-former with $N_\text{A}=N_\text{B}/4$ mixed with a sphere-former with
$N_\text{A2}=7N_\text{B}$. The Flory parameter is set to the
relatively high value of $\chi N_\text{B}=22.5$. The volume fractions of
lamella-formers as a percentage of all copolymers are (a) $5\%$, (b)
$15\%$, (c) $25\%$ and (d) $35\%$. Sphere-formers are shown with thick
lines, lamella-formers with thin lines. The hydrophobic components are
plotted with full lines, the hydrophilic components with dashed
lines, and the solvent with a dotted line.
}
\end{figure}

\begin{figure}
\includegraphics[width=0.5\linewidth]{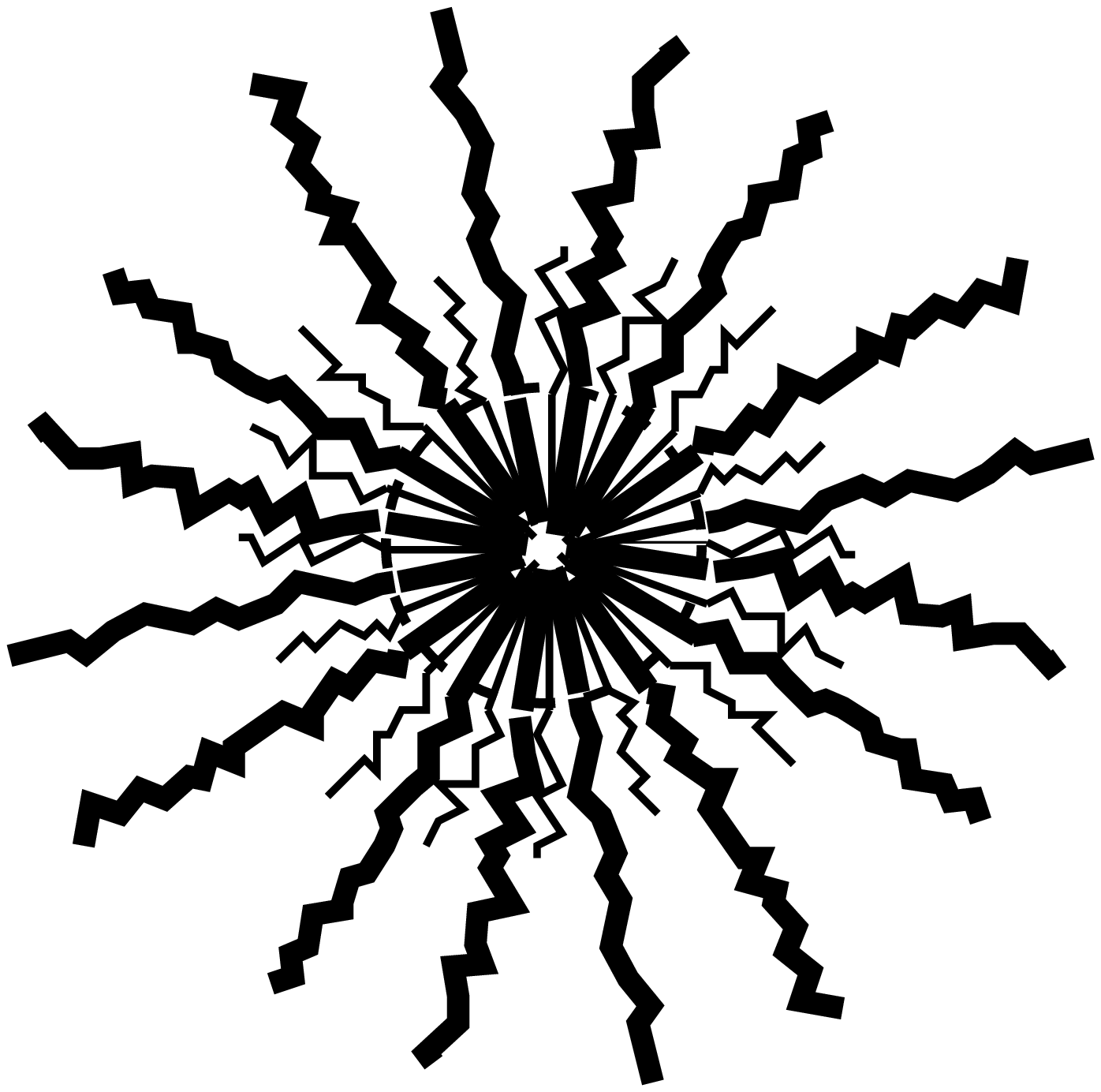}
\caption{\label{micelle_fig} Sketch of a simple mixed micelle. Sphere-formers are shown with thick
lines, lamella-formers with thin lines. The hydrophobic components are
plotted with straight lines, the hydrophilic components with zig-zag
lines, and the boundary of the hydrophobic core is marked with a
dashed circle. This structure is seen for weakly mismatched copolymers
at all $\chi$ parameters considered.
}
\end{figure}

Figure \ref{conccut45_fig} shows a series of radial cuts through the density
profile of the optimum spherical aggregate as the volume fraction of
lamella-former is increased from $5\%$ to $35\%$ of all copolymers in steps of
$10\%$. For the lowest of these lamella-former concentrations (Figure \ref{conccut45_fig}a), the
sphere-formers and lamella-formers are homogeneously mixed in the
core, and a simple mixed micelle is formed (see the sketch in Figure
\ref{micelle_fig}). This structure is little different from that
which would be formed in a system of pure sphere-forming amphiphiles:
the concentration of lamella-formers is not yet sufficiently large to
have a strong effect on the micelle morphology. Indeed, for such low
lamella-former concentrations, mixed micelles may not be present
\cite{shim}, with pure aggregates of the two species forming instead.

\begin{figure}
\includegraphics[width=0.5\linewidth]{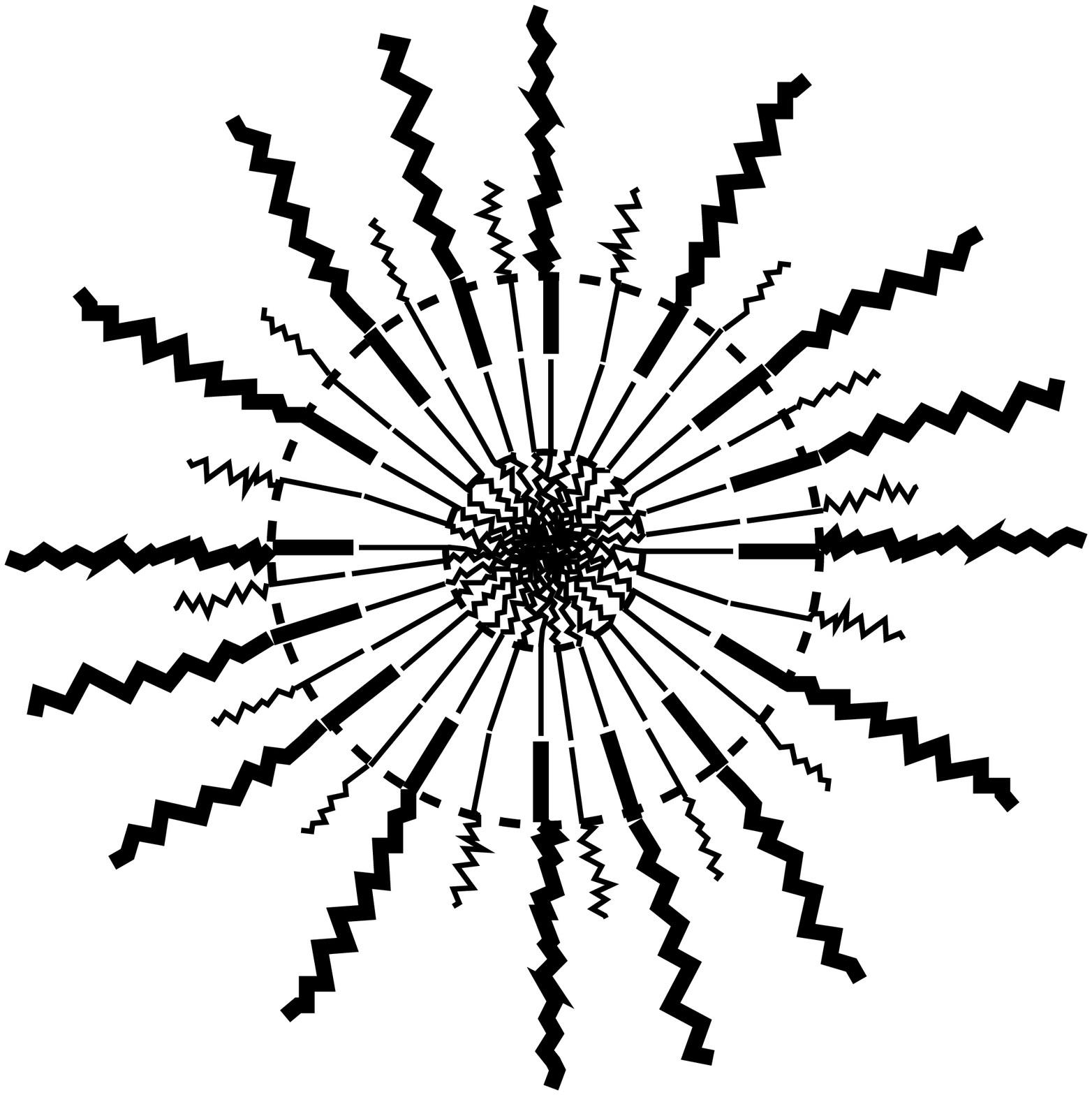}
\caption{\label{encapsulated_fig} Sketch of a complex ABA mixed micelle
  with a hydrophilic A inner core, a hydrophobic B outer core and a
  hydrophilic A corona. Sphere-formers are shown with thick
lines, lamella-formers with thin lines. The hydrophobic components are
plotted with straight lines, the hydrophilic components with zig-zag
lines, and the boundaries of the two cores are marked with
dashed circles. This structure is seen for larger $\chi$ parameters.
}
\end{figure}

However, as more lamella-forming molecules are added (Figure \ref{conccut45_fig}b), their
influence on the core composition of the micelle becomes clear. A polymer with a
large hydrophobic component that naturally forms flat bilayers or even
structures of negative curvature is in
an energetically highly unfavorable state in a small, positively
curved micelle. In consequence, these
molecules segregate to the center of the aggregate, where they form a
tightly-wrapped bilayer.  This structure is sketched in Figure
\ref{encapsulated_fig}. The polymers in the inner leaflet of this
bilayer are in a more favorable negative curvature state, with their hydrophilic
components pointing in towards the center of the micelle. Those in the
outer leaflet are also in a more favorable state than at lower
lamella-former concentrations: they are no longer
forced into the core of a compact micelle, but sit in a shell on the
outside of the new core region. Here, they are mixed with the sphere-formers,
which can no longer form their preferred simple micelle structure, but
strongly prefer the positively-curved surface of the new spherical
aggregate to its core. The new micelle therefore has an inner core of
hydrophilic A-blocks, an outer core of hydrophobic B-blocks, and a
hydrophilic A corona.

\begin{figure}
\includegraphics[width=0.5\linewidth]{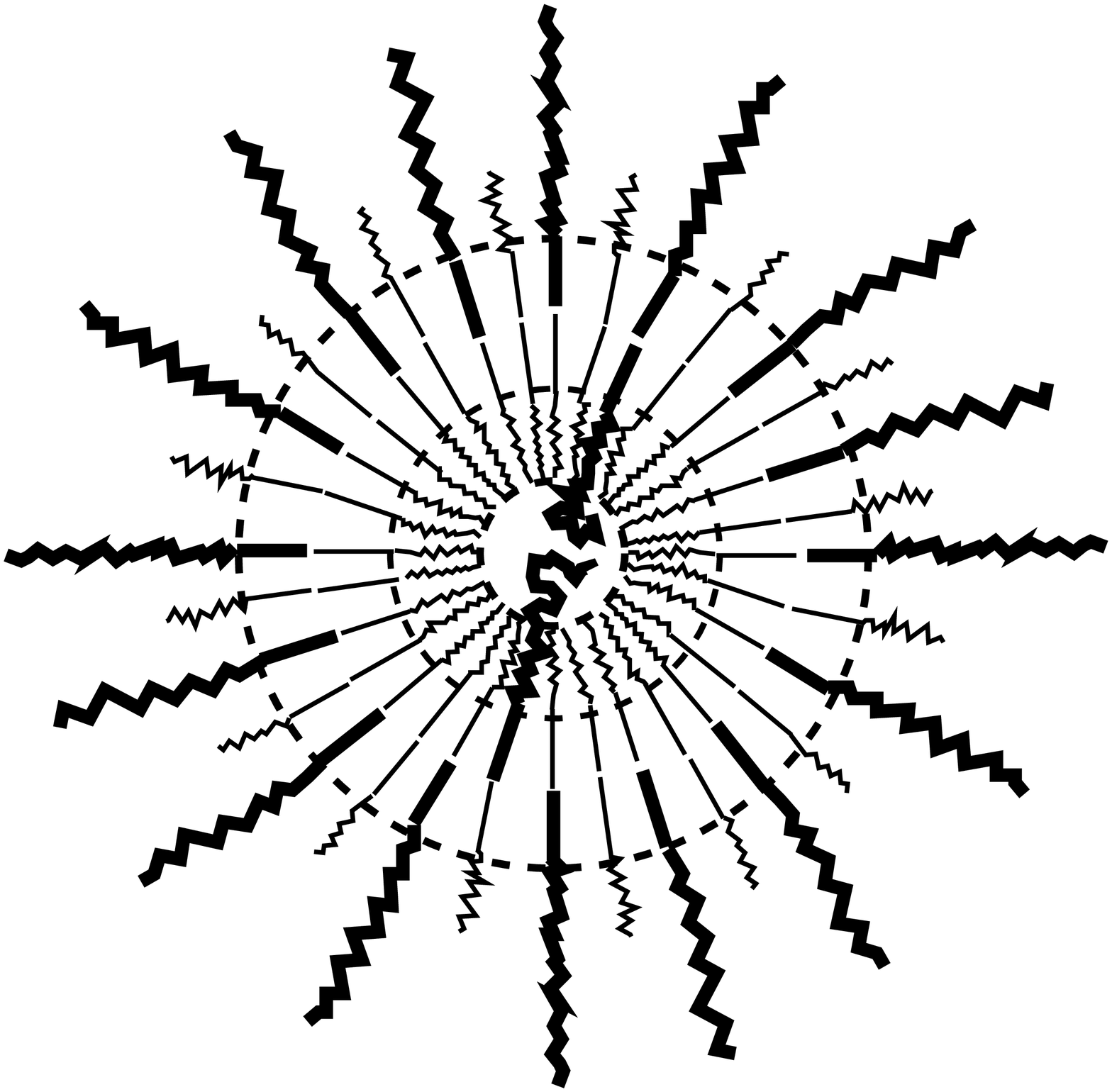}
\caption{\label{vesicle_fig} Sketch of a small vesicle
  composed of a solvent center, a layer of hydrophilic A-blocks, a layer
  of hydrophobic B-blocks and a
  hydrophilic A corona. Sphere-formers are shown with thick
lines, lamella-formers with thin lines. The hydrophobic components are
plotted with straight lines, the hydrophilic components with zig-zag
lines, and the boundaries of the various regions are marked with
dashed circles. This structure is seen for larger $\chi$ parameters.
}
\end{figure}

As the lamella-former concentration is increased still further, to
$25\%$ by volume of all copolymers (Figure \ref{conccut45_fig}c),
solvent penetrates into the core of the micelle, as the inner bilayer of
lamella-formers becomes more dominant in fixing the micelle morphology
and expands towards the planar state. This process continues in Figure
\ref{conccut45_fig}d, where $35\%$ of all copolymers are
lamella-forming. Here, a number of the sphere-formers have mixed with
the inner leaflet of the lamella-former bilayer, meaning that the overall
structure has the form of a (very) small bilayer vesicle (see the
sketch in Figure \ref{vesicle_fig}). Small vesicles with
a preferred radius have indeed been seen in experiments on mixtures of
sphere- and lamella-forming amphiphiles \cite{li,zidovska}. Their existence has also been
predicted in recent lattice SCFT calculations by Li et al.\
\cite{li2}. Our current work considers a different
region of parameter space to these lattice-based calculations, which
focus on weakly mismatched amphiphiles and (usually) higher lamella-former concentrations. In consequence, the mechanism by which the
preferred vesicle radius is selected appears to be rather different in
the two studies. In our work, strong segregation of the two
species occurs and the compositions of the inner and
outler leaflets of the bilayer are quite different. The small
vesicle structure forms as it accommodates both the preference of the
shorter copolymers for a bilayer structure and that of the longer,
sphere-forming amphiphiles for positively-curved
surfaces.
This is in
contrast to the results of Li et al.\ \cite{li2}, where the vesicles
are larger and the two
bilayer leaflets have similar compositions. The individual
vesicles therefore have no preferred curvature, and coexistence
with a high concentration of mixed micelles is found to be necessary
for vesicle size selection to occur. Highly-curved
mixed bilayers may also be seen as the end sections of larger tubular
vesicles \cite{safinya_rev}, and have also been investigated
using molecular dynamics simulations \cite{cooke_deserno}. Our current work gives some broad guidance as
to how the system parameters might be varied in order to encourage or
discourage the formation of these structures.

\begin{figure}
\includegraphics[width=\linewidth]{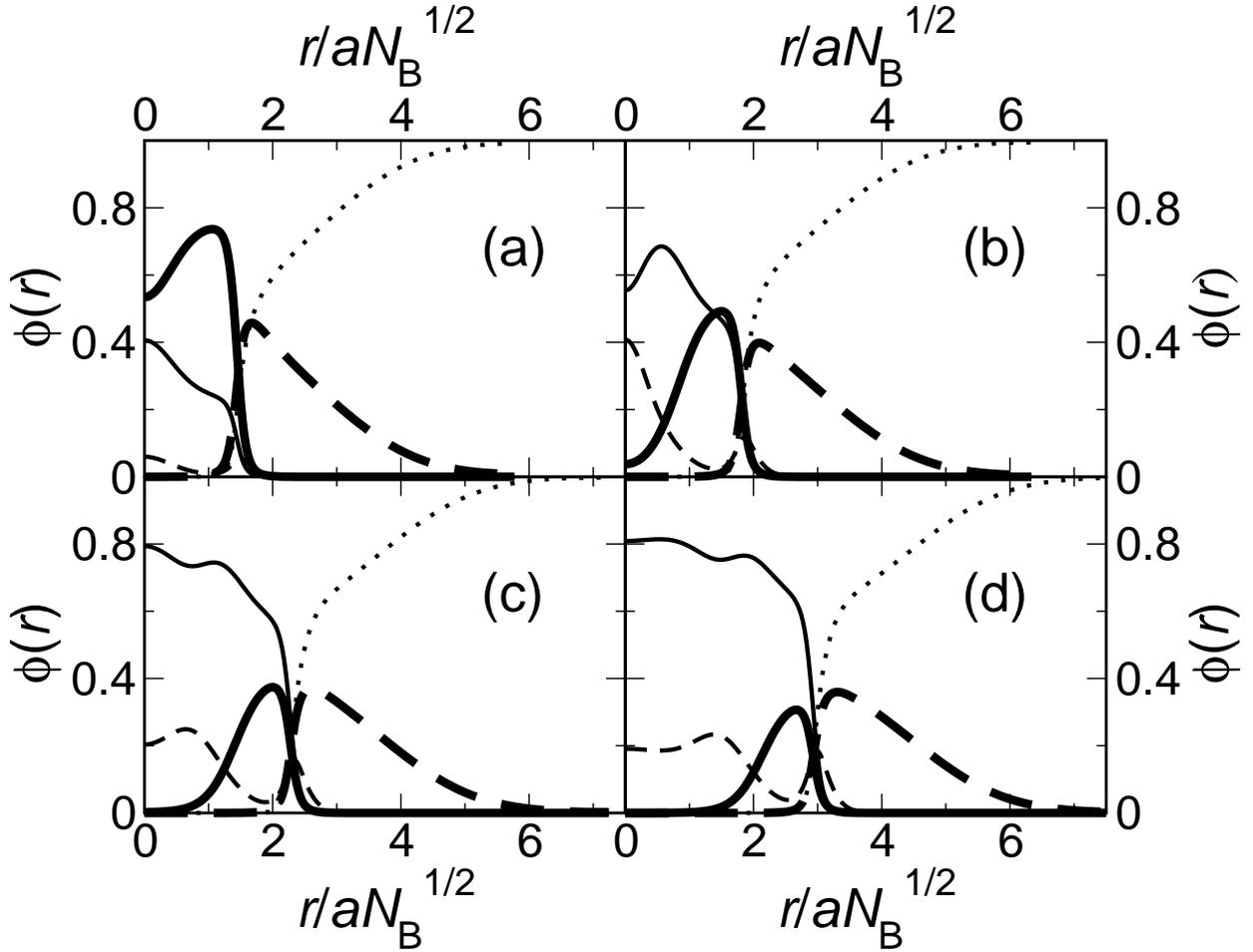}
\caption{\label{conccut30_fig} Cuts through the density profiles of the
  spherically-symmetric aggregates formed in a solution of
  lamella-former with $N_\text{A}=N_\text{B}/4$ mixed with a sphere-former with
$N_\text{A2}=7N_\text{B}$. The Flory parameter is set to the
relatively low value of $\chi N_\text{B}=15$. The volume fractions of
lamella-formers as a percentage of all copolymers are (a) $5\%$, (b)
$15\%$, (c) $25\%$ and (d) $35\%$. Sphere-formers are shown with thick
lines, lamella-formers with thin lines. The hydrophobic components are
plotted with full lines, the hydrophilic components with dashed
lines, and the solvent with a dotted line.
}
\end{figure}

\begin{figure}
\includegraphics[width=0.5\linewidth]{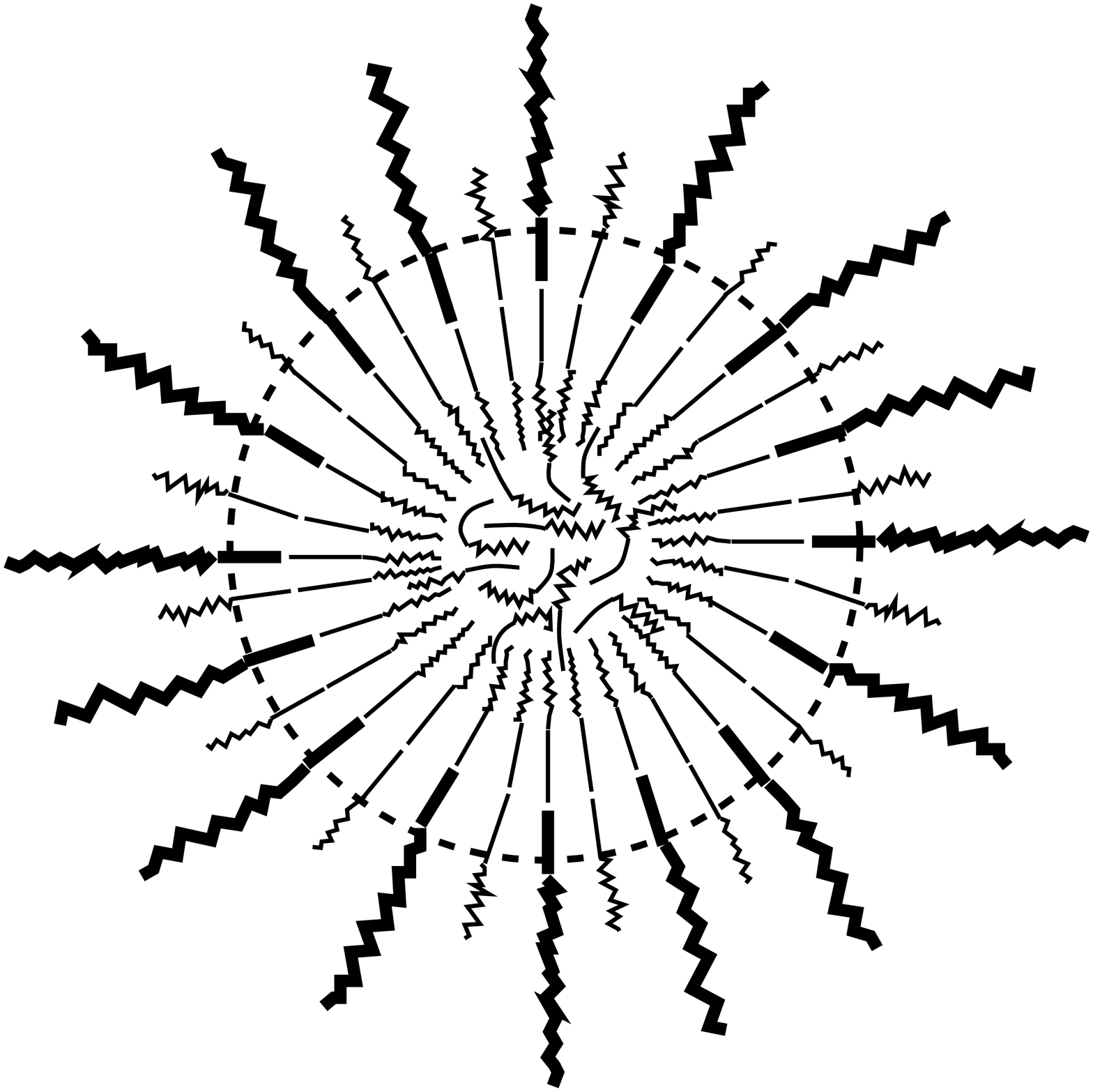}
\caption{\label{strange_fig} Sketch of a complex mixed micelle
  with a weakly structured core formed of lamella-formers and the
  hydrophobic blocks of sphere-formers. Sphere-formers are shown with thick
lines, lamella-formers with thin lines. The hydrophobic components are
plotted with straight lines, the hydrophilic components with zig-zag
lines, and the boundaries of the hydrophobic core is marked with a
dashed circle. This aggregate is seen for lower $\chi$ parameters.
}
\end{figure}

We now turn our attention to a system in which the two polymer species
have the same architectures as before ($N_\text{A}=N_\text{B}/4$ for
the lamella-former and $N_\text{A2}=7N_\text{B}$ for the sphere-former), but a significantly weaker
interaction strength of $\chi N_\text{B}=15$. The difference between
the two systems can be
seen even at only $5\%$ lamella-former (Figure \ref{conccut30_fig}a). Here, the radial
segregation of the two polymers according to their preferred
curvatures is already clearly underway. Before, it was prevented at
lower lamella-former concentrations by the energetic cost of mixing
the hydrophilic A blocks with the hydrophobic B core. As more
lamella-former is added to reach $15\%$ (Figure \ref{conccut30_fig}b), the behavior of the
two systems diverges still further. In the system with stronger
repulsive interactions discussed earlier, the A and B blocks demix in the core, leading
to the ABA structure seen in Figure \ref{conccut45_fig}b and sketched
in Figure
\ref{encapsulated_fig}. In the current system, although some demixing
does indeed occur (Figure \ref{conccut30_fig}b), the effect is much weaker,
and A- and B-rich regions can no longer be clearly separated. This
structure is sketched in Figure
\ref{strange_fig}. For even larger concentrations of lamella-former of
$25\%$ and $35\%$
(Figure \ref{conccut30_fig}c and d), the small vesicle structure seen before is
absent. Instead, since the A and B blocks may mix much more freely
than before, the core of the micelle is formed of a nearly homogeneous
melt of lamella-former.

\begin{figure}
\includegraphics[width=\linewidth]{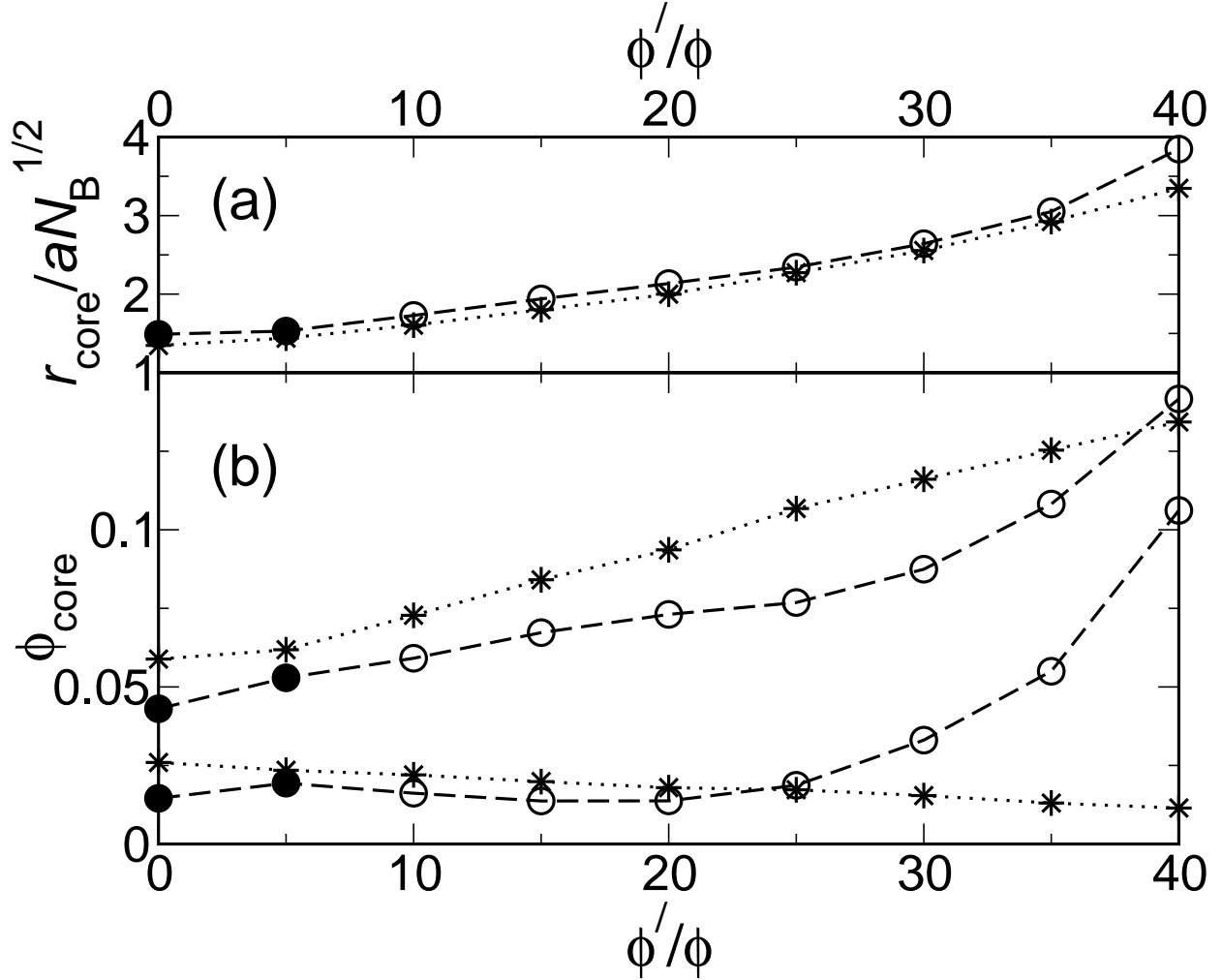}
\caption{\label{radius_comp_conc_fig} Core radius and composition of the
  spherically-symmetric aggregates formed in a solution of
  lamella-former with $N_\text{A}=N_\text{B}/4$ mixed with a sphere-former with
$N_\text{A2}=7N_\text{B}$. (a) Core radius as a function of the ratio $\phi'/\phi$ of the volume fraction of
lamella-formers $\phi'$ to the total volume fraction of copolymers
$\phi$. Points corresponding to simple micelles (Figure \ref{micelle_fig}) are marked by closed
circles, ABA aggregates (Figure \ref{encapsulated_fig}) or small vesicles
(Figure \ref{vesicle_fig}) are marked with open
circles, and weakly-structured aggregates (Figure \ref{strange_fig}) by asterisks. The data for
the system with a Flory parameter of $\chi N_\text{B}=15$ are connected with
dotted lines; those for the system with $\chi N_\text{B}=22.5$ by
dashed lines. (b) Fraction of the core that is composed of A-blocks for each of the two
systems (upper two curves), and fraction of the core
that is composed of homopolymer solvent (lower two curves).
}
\end{figure}

To make our study of the concentration dependence of the binary system
more quantitative, we calculate the core radius and composition
as a function of the ratio $\phi'/\phi$ of the volume fraction of
lamella-formers $\phi'$ to the total volume fraction of copolymers $\phi$. We define the core boundary as the radius at which
the volume fraction of hydrophobic blocks is equal to $0.5$, and plot
this quantity in Figure \ref{radius_comp_conc_fig}a for both
systems considered above. The core radii of both species grow steadily and almost identically as
the lamella-former concentration is increased. In the case of the
system with $\chi N_\text{B}=22.5$, the growth is associated with the
expansion of the micelle to form a small vesicle, while in the system with weaker interactions
($\chi N_\text{B}=15$), it arises from the fact that the core is gradually filling with
lamella-former. The only appreciable
difference in radius is seen in the final point, where $40\%$ of all copolymers are
lamella-forming. Here, the radius of the $\chi N_\text{B}=22.5$ system
has begun to grow more rapidly, as the system moves towards the planar
bilayer state. For lamella-former fractions greater than $40\%$, the
influence of the sphere-forming copolymers is weak, and we
were no longer able to find free-energy minima corresponding to small
spherical aggregates.

The contrast between the high and low interaction strength systems emerges more clearly if we
consider the amounts of the different species in the core defined
above. The upper two curves in Figure
\ref{radius_comp_conc_fig}b show the fraction of the core that is composed of A-blocks for each of the two
systems, while the lower two curves show the fraction of the core
that is composed of homopolymer solvent. In the $\chi N_\text{B}=15$
system, the amount of A-block in the core grows steadily as the amount
of lamella-former is increased, as in this case the lamella-forming copolymer is
simply encapsulated in the center of the micelle. For low
lamella-former concentrations, the $\chi N_\text{B}=22.5$ system forms
simple micelles with a clear interface in between the core and corona,
and so has less A-block in the core than does the $\chi
N_\text{B}=15$ system, where some mixing of A and B blocks occurs in
the core. In contrast, for larger amounts of lamella-forming
copolymer, the fraction of A-block in the more strongly-interacting
system grows more and more rapidly as the
preferred aggregate changes from a closed micelle to an open vesicle.

The clearest difference between the two systems is seen
in the behavior of the amount of solvent in the core as the
lamella-former concentration is increased (see the lower two curves in Figure
\ref{radius_comp_conc_fig}b). In the $\chi N_\text{B}=15$
system, where the core is largely composed of copolymers, the fraction of solvent remains fairly constant at around
$0.02$-$0.03$. The fact that this value is a little higher than might
be expected from the density profiles in Figure \ref{conccut30_fig}, and also
varies slightly, can be
attributed to the fact that our simple definition of the core radius means 
that a thin shell of solvent is always included as part of the
core. In the system with stronger repulsion between the A and B
components, the core solvent fraction starts at a similar small value,
remaining close to this as the lamella-former volume fraction is
increased and the morphology of the system changes from a simple
micelle to the ABA structure shown in Figure \ref{encapsulated_fig}. However, as the
fraction of lamella-formers $\phi'/\phi$ is increased towards $40\%$, the fraction of solvent in the core
grows very quickly as the aggregate expands towards a vesicle.

We have checked a selection of these calculations for a much more dilute
system with an overall copolymer volume fraction of
$\phi\approx 1\%$, focusing in particular on those lamella-former volume fractions
where the aggregate changes from one morphology to another. Such a dilute system may be more appropriate for observation of
the small spherical structures considered in this paper, since a more concentrated solution might form larger aggregates such as worm-like micelles. We
therefore wish
to check that the form of the optimum spherical aggregates is not
strongly sensitive to concentration (although the likelihood of their
formation with respect to larger aggregates may of course depend on the concentration).

Indeed, 
little change in the form of the aggregates is
observed. The most significant difference is that, in the dilute case,
the transition between the simple micelle and the ABA structure in the
$\chi N_\text{B}=22.5$ system occurs when $\phi'/\phi$ is between
$15\%$ and $20\%$, rather than between $10\%$ and $15\%$ in the
$\phi=10\%$ system. This preference for small spherical micelles in more dilute
systems is in line with the known concentration dependence of block
copolymer solutions \cite{kinning_winey_thomas}. Similar small shifts
in the morphology transitions as the overall copolymer concentration
is varied, or no appreciable shifts at all, are observed in
all the systems considered in this paper, where the polymers
considered aggregate reasonably strongly and the free-energy
minima associated with the various micelle shapes will be relatively
sharp. This is not the case for shorter or more weakly-interacting
polymers, where the concentration dependence may be quite strong.

\subsection{Effect of interaction strength}

\begin{figure}
\includegraphics[width=\linewidth]{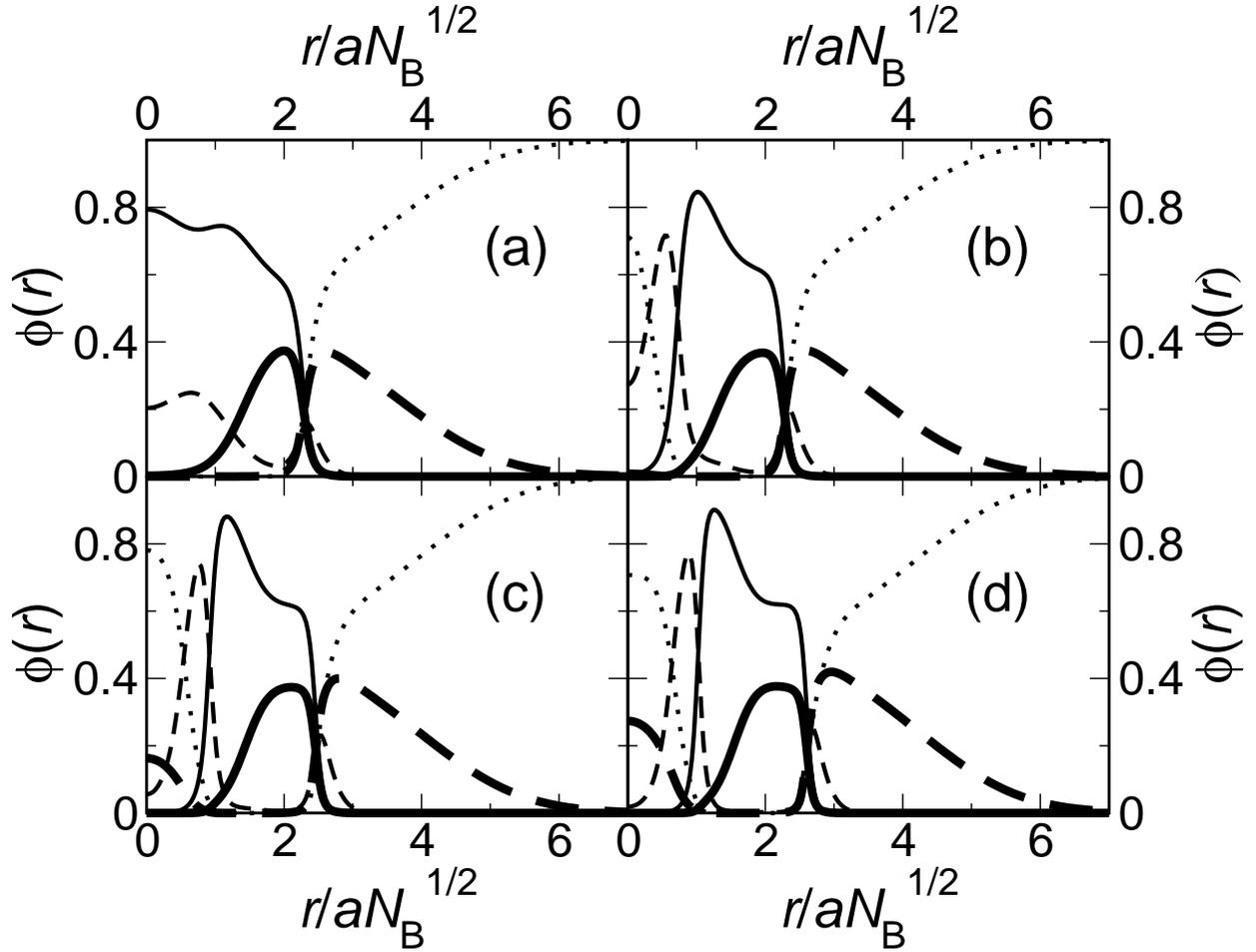}
\caption{\label{chicut_fig} Cuts through the density profiles of the
  spherically-symmetric aggregates formed in a solution of
  lamella-former with $N_\text{A}=N_\text{B}/4$ mixed with a sphere-former with
$N_\text{A2}=7N_\text{B}$. The lamella-former fraction $\phi'/\phi$ is
set to
$25\%$. The Flory parameter $\chi N_\text{B}$ is varied and takes the following values: (a) $15$, (b)
$20$, (c) $25$ and (d) $30$. Sphere-formers are shown with thick
lines, lamella-formers with thin lines. The hydrophobic components are
plotted with full lines, the hydrophilic components with dashed
lines, and the solvent with a dotted line.}
\end{figure}

In the results discussed in the preceding subsection, we found a clear
contrast in phenomenology between two systems with different levels of
repulsion between their hydrophilic and hydrophobic components. To
investigate this effect in more detail, we take a system with the same
copolymer architectures as considered above ($N_\text{A}=N_\text{B}/4$ for
the lamella-former and $N_\text{A2}=7N_\text{B}$ for the
sphere-former), fix the lamella-former fraction $\phi'/\phi$ to
$25\%$, and vary $\chi N_\text{B}$. Figure \ref{chicut_fig} shows a series
of cuts through the density profiles of the small spherical aggregates
formed as $\chi N_\text{B}$ is increased from $15$ to $30$ in steps of
$5$. For the smallest of these values, shown in Figure
\ref{chicut_fig}a, we find a weakly-structured aggregate of the kind
shown in Figure \ref{conccut30_fig}. Very similar results are also found for
the even smaller value of $\chi N_\text{B}=12.5$. Below this value, no
self-assembly takes place.

As $\chi N_\text{B}$ is increased to
$20$ (Figure \ref{chicut_fig}b), the segregation between hydrophilic and hydrophobic
blocks has become much stronger, and the ABA aggregate of Figure
\ref{encapsulated_fig} is seen, with some penetration of solvent into
the core region. On further increase of $\chi N_\text{B}$, the
boundaries between the various layers become sharper and
sharper as the repulsive interaction between the A and B blocks
increases in strength. The radius also increases, and the aggregate
opens out to a small vesicle. Finally, in Figure \ref{chicut_fig}d, where $\chi
N_\text{B}=30$, the layers are very clearly defined, and the
calculated density profile
resembles very closely the vesicle sketched in Figure \ref{vesicle_fig}.

\subsection{Effect of lamella-forming architecture}

In the two previous subsections, we considered the effect of blending
two strongly mismatched copolymers, to demonstrate the effects of
segregation on the micelle morphologies as clearly as possible. We now
investigate how the micelle shapes change as the mismatch between the
two polymers is decreased. Specifically, the hydrophilic block
size of
the lamella-former is gradually increased from the small value
$N_\text{A}=N_\text{B}/4$ used in the preceding calculations until we
reach the symmetric molecule with
$N_\text{A}=N_\text{B}$. The same sphere-former architecture as before
is used, with $N_\text{A2}=7N_\text{B}$. Since we wish to focus
specifically on the
effects of the lamella-former, we use a slightly higher fraction of
these molecules than in the preceding section, and set
$\phi'/\phi=33.3\%$.

\begin{figure}
\includegraphics[width=\linewidth]{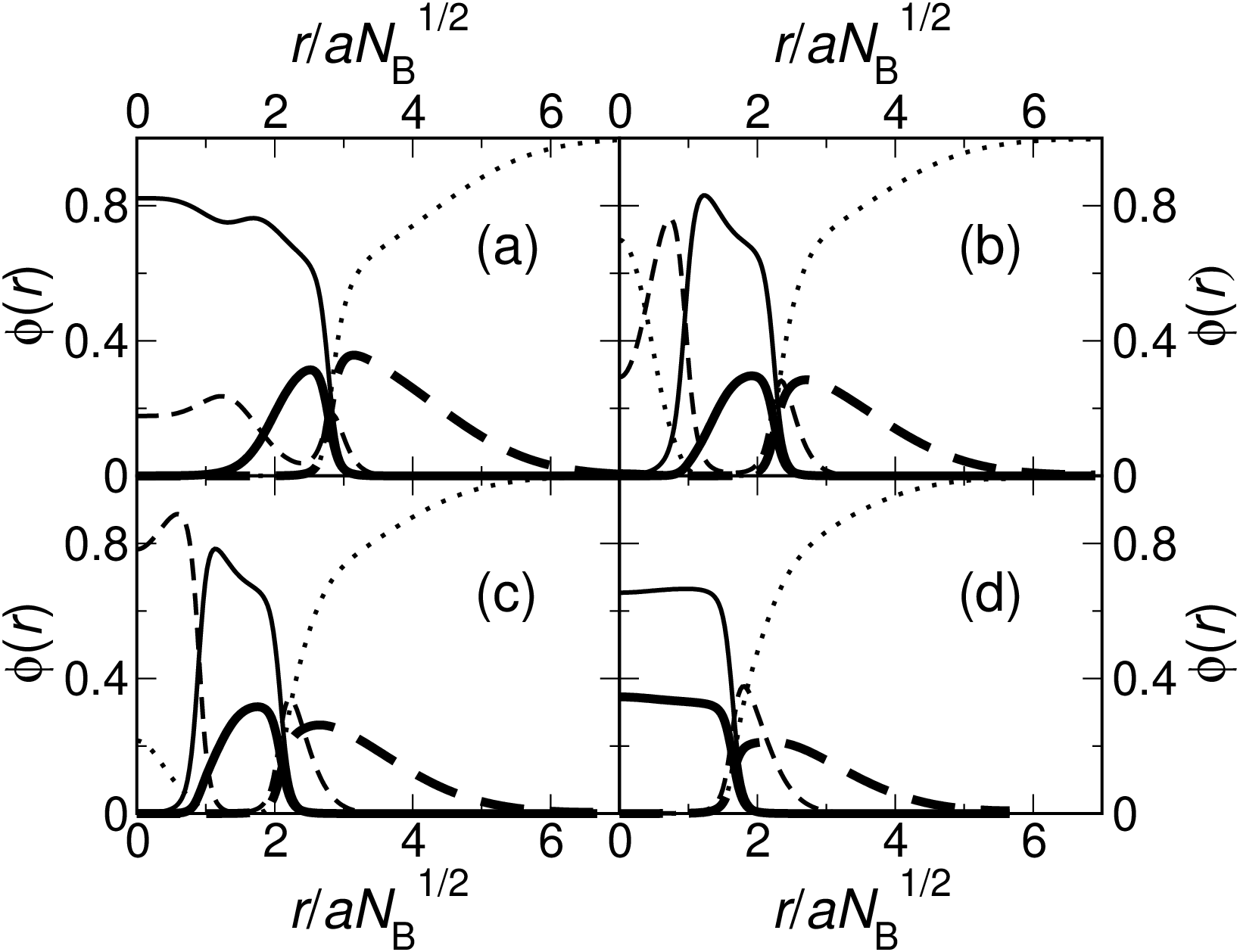}
\caption{\label{lamcut30_fig} Cuts through the density profiles of the
  spherically-symmetric aggregates formed in a solution of sphere-former with
$N_\text{A2}=7N_\text{B}$ and lamella-formers of varying architecture. The Flory parameter is set to the
relatively low value of $\chi N_\text{B}=15$. The lamella-former fraction $\phi'/\phi$ is
set to
$33.3\%$. The hydrophilic block lengths of the lamella-forming
molecules are (a), $N_\text{A}=N_\text{B}/4$ (b) $N_\text{A}=3N_\text{B}/7$, (c) $N_\text{A}=2N_\text{B}/3$ and (d) $N_\text{A}=N_\text{B}$. Sphere-formers are shown with thick
lines, lamella-formers with thin lines. The hydrophobic components are
plotted with full lines, the hydrophilic components with dashed
lines, and the solvent with a dotted line.
}
\end{figure}

To begin, we set $\chi N_\text{B}=15$, and first
consider the strong lamella-former with
$N_\text{A}=N_\text{B}/4$. This is the system of Figure \ref{conccut30_fig},
and, as there, we find large micelles with a weakly-structured core
(see Figure \ref{lamcut30_fig} a). A sharp change is observed when the length of the hydrophilic
component of the lamella-forming copolymer is increased to
$N_\text{A}=3N_\text{B}/7$ (Figure \ref{lamcut30_fig}b). Here, the AB
interfaces within the core become well-defined, and the ABA aggregate
sketched in Figure \ref{encapsulated_fig} is seen, with some penetration of
solvent into the core. This indicates that the formation of the
weakly-structured aggregate seen in Figure \ref{lamcut30_fig}a requires not only
a relatively small $\chi$ parameter between the hydrophilic and
hydrophobic blocks, but also a short hydrophilic block of the lamella
former. If the length of this block is increased, the effective
strength $\chi N$ of the
interaction between the A and B blocks of the lamella-former becomes
larger \cite{jones_book} and the two blocks can segregate within the core.

As the length of the hydrophilic block of the lamella-former is
increased still further, to $N_\text{A}=2N_\text{B}/3$, the mismatch
between the two species decreases and the aggregate shows the first
signs of approaching the small micelle favored by the
sphere-formers. Specifically, the core radius decreases slightly, and
the solvent begins to be expelled from the center of the micelle
(Figure \ref{lamcut30_fig}c). This process is complete in Figure
\ref{lamcut30_fig}d, where the lamella former is symmetric and
$N_\text{A}=N_\text{B}$. Here, a simple mixed micelle is formed, with
no segregation of the two species.

\begin{figure}
\includegraphics[width=\linewidth]{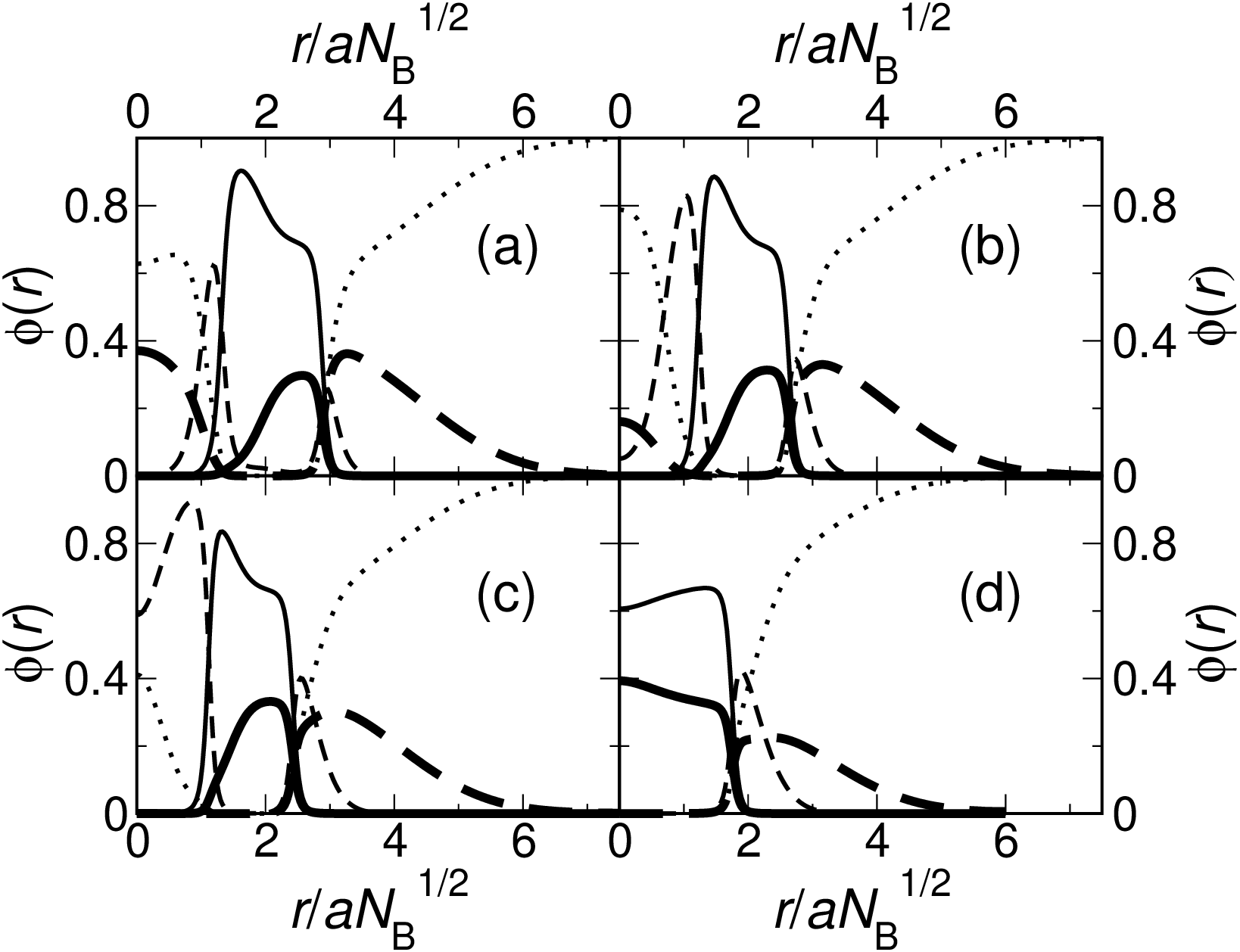}
\caption{\label{lamcut45_fig} Cuts through the density profiles of the
  spherically-symmetric aggregates formed in a solution of sphere-former with
$N_\text{A2}=7N_\text{B}$ and lamella-formers of varying architecture. The Flory parameter is set to the
relatively high value of $\chi N_\text{B}=22.5$. The lamella-former fraction $\phi'/\phi$ is
set to
$33.3\%$. The hydrophilic block lengths of the lamella-forming
molecules are (a), $N_\text{A}=N_\text{B}/4$ (b) $N_\text{A}=3N_\text{B}/7$, (c) $N_\text{A}=2N_\text{B}/3$ and (d) $N_\text{A}=N_\text{B}$. Sphere-formers are shown with thick
lines, lamella-formers with thin lines. The hydrophobic components are
plotted with full lines, the hydrophilic components with dashed
lines, and the solvent with a dotted line.
}
\end{figure}

We now consider a system with the same sequence of polymer architectures as above,
but with a larger repulsive interaction strength $\chi N_\text{B}=22.5$ between the A and B blocks. For the shortest lamella-former, with
$N_\text{A}=N_\text{B}/4$, a small vesicle forms (Figure \ref{lamcut45_fig}a), in contrast to the weakly-structured
aggregate seen for this architecture for the smaller value of $\chi
N_\text{B}$ (Figure \ref{lamcut30_fig} a). As the hydrophilic block length of
the lamella-former is increased, and the degree of mismatch between
the two copolymer species lessens, the vesicle contracts (Figure \ref{lamcut45_fig}b and
c) until a simple mixed micelle is formed (Figure \ref{lamcut45_fig}d). 

\begin{figure}
\includegraphics[width=\linewidth]{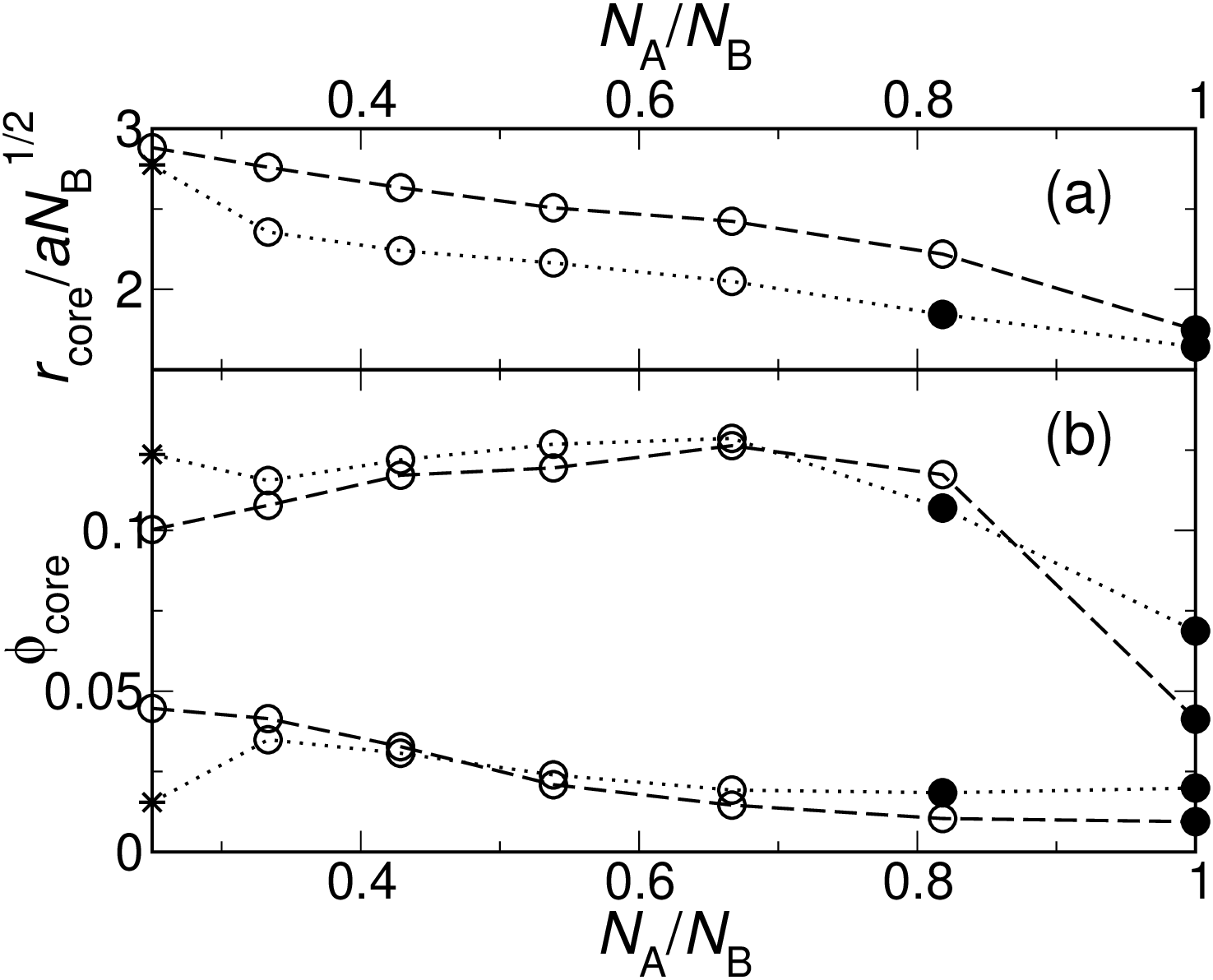}
\caption{\label{radius_comp_lam_fig} Core radius and composition of the
  spherically-symmetric aggregates formed in a solution of sphere-former with
$N_\text{A2}=7N_\text{B}$ mixed with lamella-formers of various
architectures. The lamella-former fraction $\phi'/\phi$ is
set to
$33.3\%$. (a) Core radius as a function of the
hydrophilic block length. Points corresponding to simple micelles
(Figure \ref{micelle_fig}) are marked by closed
circles, ABA aggregates (Figure \ref{encapsulated_fig}) or small vesicles
(Figure \ref{vesicle_fig}) are marked with open
circles, and weakly-structured aggregates (Figure \ref{strange_fig}) by asterisks. The data for
the system with a Flory parameter of $\chi N_\text{B}=15$ are connected with
dotted lines; those for the system with $\chi N_\text{B}=22.5$ by
dashed lines. (b) Fraction of the core that is composed of A-blocks for each of the two
systems (upper two curves), and fraction of the core
that is composed of homopolymer solvent (lower two curves). 
}
\end{figure}

As in our discussion of the dependence of the micelle morphology on
lamella-former concentration, we now plot the aggregate core radii and
composition as a function of lamella-former architecture for the two
systems studied in this section. In Figure \ref{radius_comp_lam_fig}a,
we show the decrease of the micelle radius as the length of the
lamella-former hydrophilic block is increased. The lower line shows
the behavior of the radius of the system with the smaller Flory
parameter $\chi N_\text{B}=15$. The sharpest change in the radius
occurs between the first two points, when the aggregate changes from
the weakly segregated structure plotted in Figure \ref{lamcut30_fig}a
to an ABA micelle with the form shown in Figure \ref{lamcut30_fig}b. This latter
structure then gradually contracts as the lamella-former is lengthened
until we arrive at the simple mixed micelle plotted in Figure
\ref{lamcut30_fig}d. This steady contraction with increasing lamella-former
length is also seen in the
$\chi N_\text{B}=22.5$ system. The small vesicle/ABA morphology is
especially robust here, being formed not only for the short
lamella-former (which formed a weakly-structured aggregate for $\chi
N_\text{B}=15$) but also for all other lamella-formers
apart from the longest with $N_\text{A}=N_\text{B}$.

The fraction of the core that is composed of A-blocks displays
especially interesting behavior as the lamella-former hydrophilic
block length is varied. In the system with the smaller Flory parameter
$\chi N_\text{B}=15$, the A-block fraction has a rather high
value of around $0.125$ for the short lamella-formers with
$N_\text{A}=N_\text{B}/4$. This is because the system forms a
weakly-structured aggregate (Figure \ref{lamcut30_fig}a) here, with a core composed of
lamella-forming copolymers (Figure \ref{strange_fig}). As the hydrophilic
blocks of the lamella-formers
are lengthened, solvent enters into the core and the A-block fraction
falls slightly. Further increase of the lamella-former hydrophilic
block length causes the fraction of A-blocks in the core to rise steadily. The reason
for this is that, as the
aggregate shrinks and solvent is slowly expelled from the center of
the aggregate, the
amount of A-blocks changes relatively little. These blocks therefore
come to constitute a larger fraction of the core. As the
lamella-former A-blocks are lengthened still further, we observe a
sharp drop in the fraction of hydrophilic material in the core, as the
system contracts to form a simple mixed micelle.

Some aspects of this behavior are also seen in the $\chi
N_\text{B}=22.5$ solution. Here, the weakly-structured aggregate of Figure
\ref{lamcut30_fig}a is not present, and the system forms a small
vesicle in the case of the shortest lamella-formers. As the A-block
length of
these molecules is increased, the amount of solvent in the core of the
aggregate decreases, and the fraction of A-blocks in the core rises,
as in the case of the $\chi N_\text{B}=15$ system. For the largest
A-block lengths studied, the fraction of hydrophilic material in the
core is much smaller, as the system has formed a mixed micelle with a
predominantly hydrophobic core.

To reinforce the above arguments, we also show in Figure \ref{radius_comp_lam_fig}b the volume fraction of
the core that is composed of solvent as a function of lamella-former
hydrophilic block length. For the solution with $\chi
N_\text{B}=15$, the core solvent fraction initially rises as the weakly segregated
structure is replaced by an ABA structure with some solvent in the
core. It then falls gradually as the aggregate contracts to form a mixed
micelle. A similar steady fall is observed in the more strongly
segregated $\chi N_\text{B}=22.5$ system, as the small open vesicle
observed for small $N_\text{A}$ closes to form a micelle.

\subsection{Effect of sphere-forming architecture}

\begin{figure}
\includegraphics[width=\linewidth]{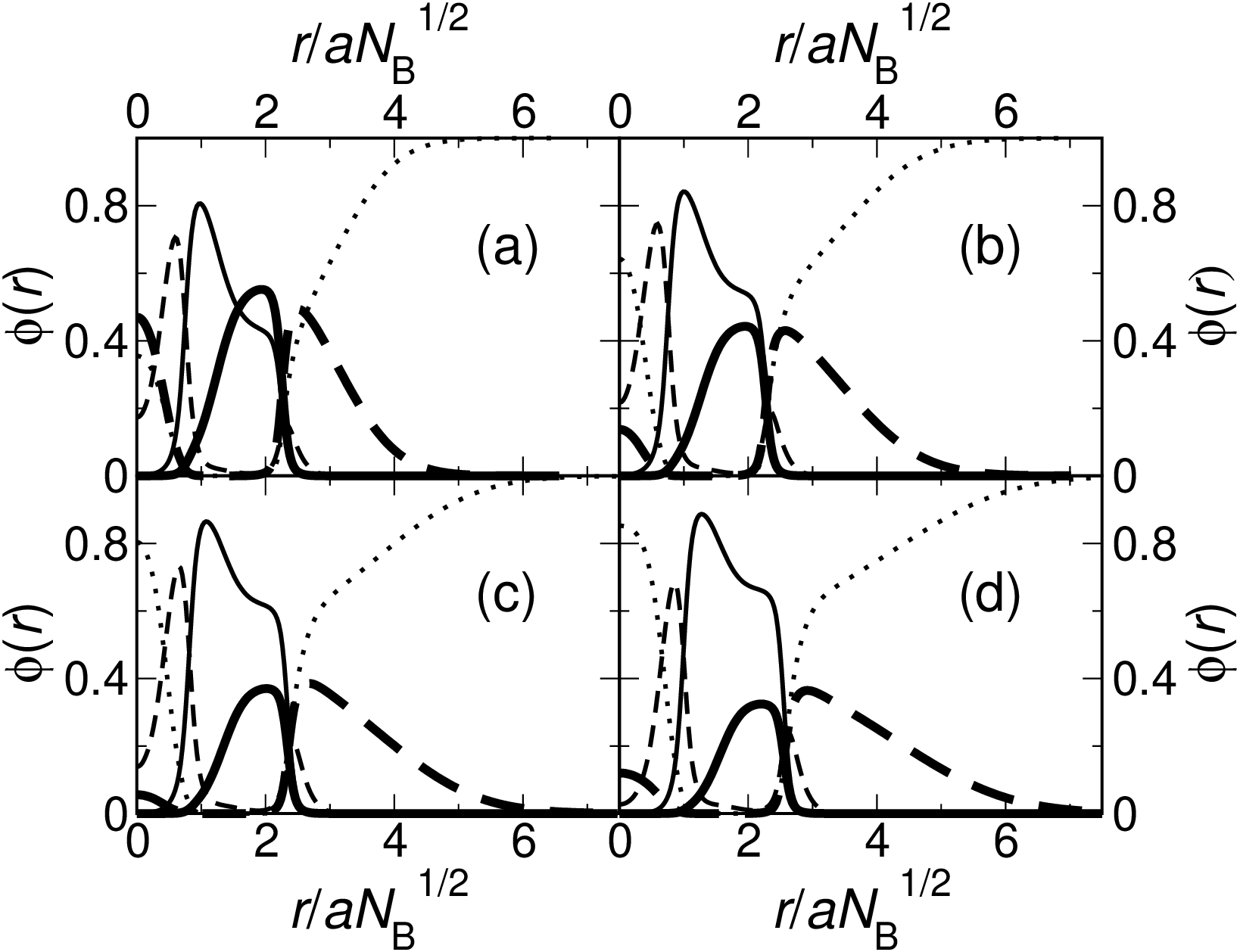}
\caption{\label{spherecut45_fig} Cuts through the density profiles of the
  spherically-symmetric aggregates formed in a solution of lamella-former with
$N_\text{A}=N_\text{B}/4$ and sphere-formers of varying architecture. The Flory parameter is set to the
relatively high value of $\chi N_\text{A}=22.5$. The lamella-former fraction $\phi'/\phi$ is
set to
$25\%$. The hydrophilic lengths of the sphere-forming
molecules are (a), $N_\text{A2}=3N_\text{B}$ (b) $N_\text{A2}=5N_\text{B}$, (c) $N_\text{A2}=7N_\text{B}$ and (d) $N_\text{A2}=9N_\text{B}$. Sphere-formers are shown with thick
lines, lamella-formers with thin lines. The hydrophobic components are
plotted with full lines, the hydrophilic components with dashed
lines, and the solvent with a dotted line.
}
\end{figure}

To conclude the scan of our system's parameter space, we now
investigate the effect of the architecture of the sphere-former on the
morphology of the aggregates. In the above results, we focused on
strongly mismatched copolymers and so used a highly asymmetric
sphere-former with $N_\text{A2}=7N_\text{B}$. We now vary the length of
the hydrophilic block of the sphere-forming copolymer over a wide
range, starting from a short molecule with $N_\text{A2}=3N_\text{B}$
and increasing the number of A monomers until
$N_\text{A2}=9N_\text{B}$. The architecture of the lamella-former is
fixed, with $N_\text{A}=N_\text{B}/4$, and, as in all the above cases,
the total copolymer volume fraction is kept constant at $10\%$. Three
quarters of these copolymers are sphere-forming, so that
$\phi'/\phi=25\%$. As in our studies of the effect of copolymer concentration and
lamella-former architecture, we consider two values of the Flory
parameter: $\chi N_\text{B}=22.5$ and $\chi N_\text{B}=15$.

In Figure \ref{spherecut45_fig}a to d, we show cuts through the density profiles of
the optimum aggregates formed when
$N_\text{A2}=3N_\text{B}$, $5N_\text{B}$, $7N_\text{B}$ and
$9N_\text{B}$ for $\chi N_\text{B}=15$. Despite the wide variation in the number of hydrophilic
monomers, roughly similar small vesicle structures are formed in all
cases, with particularly little change in morphology being observed between
$N_\text{A2}=5N_\text{B}$ and $9N_\text{B}$. Provided the two copolymer species are sufficiently
strongly mismatched to segregate within the aggregate, there is indeed
no reason to suspect that increasing the sphere-former A-block length
further should cause major qualitative changes to the form of the
aggregate, as the sphere-formers have already reached the outer
surface and can move no further. In fact, the differences between the
four panels of Figure \ref{spherecut45_fig} can be attributed mainly to the
fact that increasing the length of the sphere-former hydrophilic
block at constant $\phi'/\phi$ gradually reduces the amount of sphere-former
hydrophobic block, with the result that the hydrophobic
core becomes more and more dominated by the lamella-former. In
consequence, the core radius of the aggregate increases somewhat, as
the lamella-formers push outwards towards their preferred flat state.

\begin{figure}
\includegraphics[width=\linewidth]{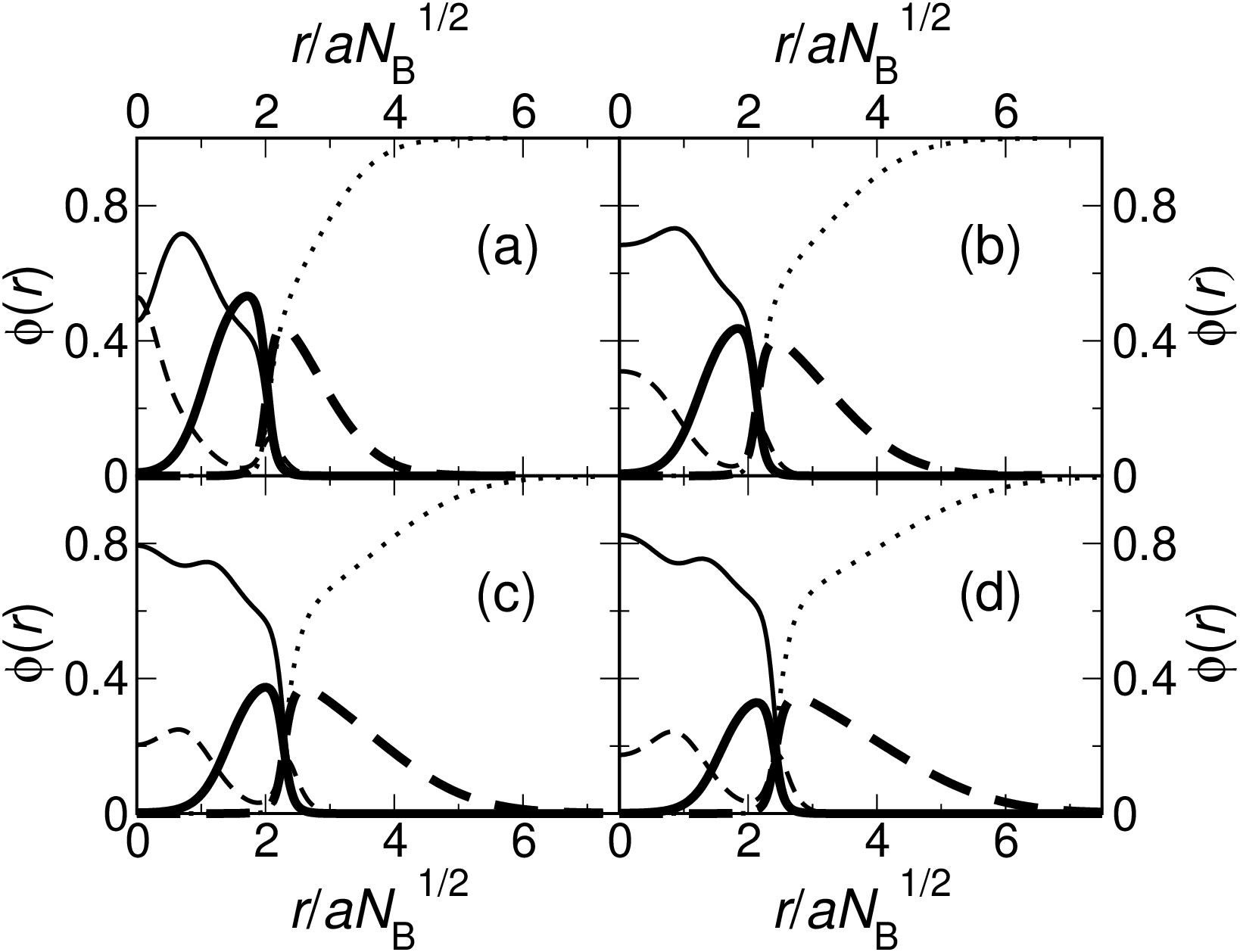}
\caption{\label{spherecut30_fig} Cuts through the density profiles of the
  spherically-symmetric aggregates formed in a solution of lamella-former with
$N_\text{A}=N_\text{B}/4$ and sphere-formers of varying architecture. The Flory parameter is set to the
relatively low value of $\chi N_\text{B}=15$. The lamella-former fraction $\phi'/\phi$ is
set to
$25\%$. The hydrophilic lengths of the sphere-forming
molecules are (a), $N_\text{A2}=3N_\text{B}$ (b) $N_\text{A2}=5N_\text{B}$, (c) $N_\text{A2}=7N_\text{B}$ and (d) $N_\text{A2}=9N_\text{B}$. Sphere-formers are shown with thick
lines, lamella-formers with thin lines. The hydrophobic components are
plotted with full lines, the hydrophilic components with dashed
lines, and the solvent with a dotted line.
}
\end{figure}

The dependence of the aggregate shape on the hydrophilic block length of the sphere-formers
is similarly weak for the smaller Flory parameter $\chi
N_\text{B}=15$. Here, aggregates with the same basic form of an outer layer of
sphere-forming copolymers encapsulating a weakly-structured core of
lamella-formers are seen for $N_\text{A}=3N_\text{B}$, $5N_\text{B}$, $7N_\text{B}$ and
$9N_\text{B}$ (Figure \ref{spherecut30_fig}a-d). As in the $\chi
N_\text{B}=22.5$ case, rather little difference in morphology can be seen as the sphere-former
hydrophilic block length is increased from $5N_\text{B}$ to
$9N_\text{B}$, save for a fall in the density of the outer
sphere-former layer of the core and a slow
growth in the core radius. The explanation for these changes is also
the same as in the system with a higher Flory parameter. Specifically, the
gradual fall in the amount of sphere-former hydrophobic block means
that the core becomes predominantly composed of lamella-forming
copolymers, which also causes it to swell.

\begin{figure}
\includegraphics[width=\linewidth]{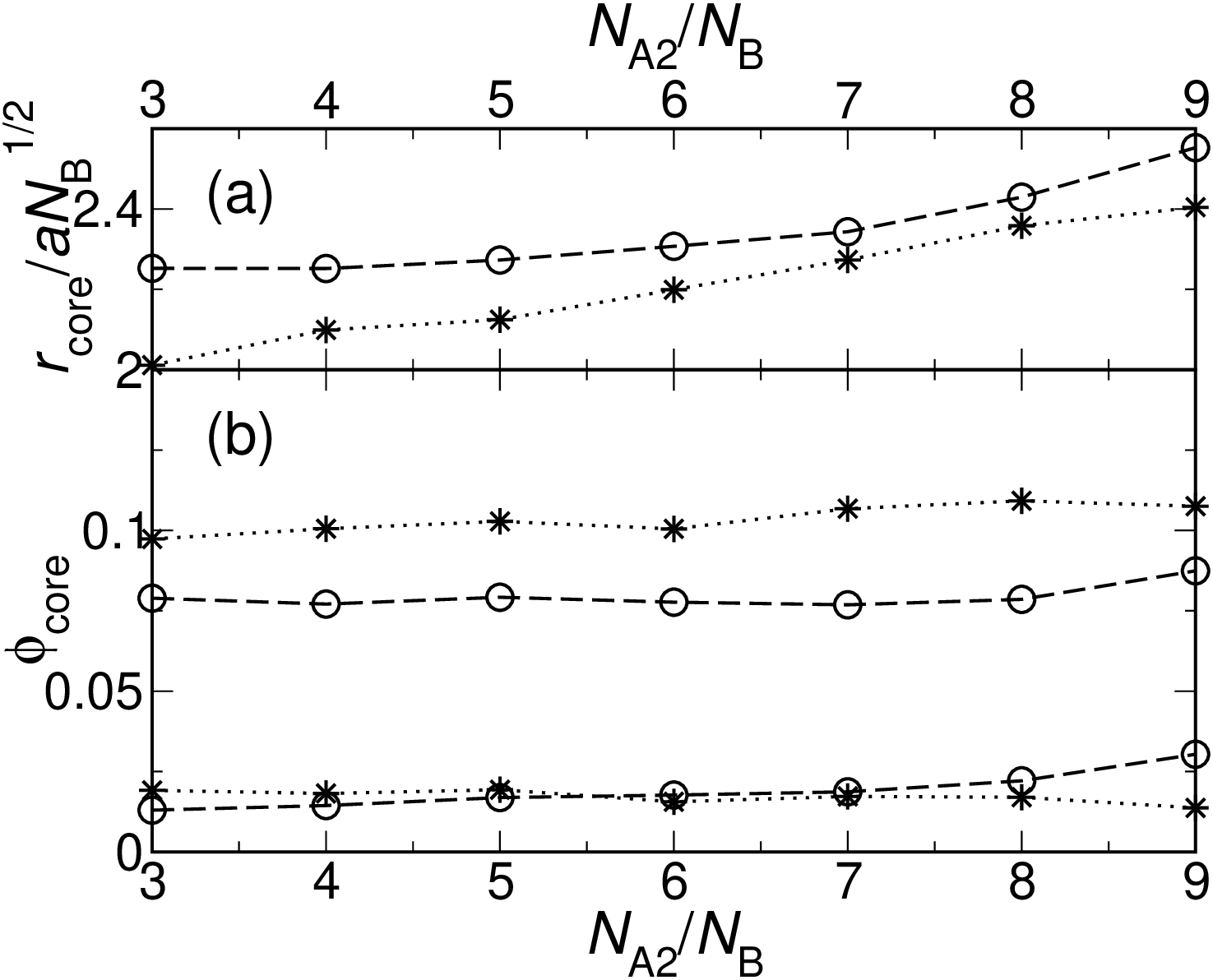}
\caption{\label{radius_comp_sphere_fig} Core radius and composition of the
  spherically-symmetric aggregates formed in a solution of lamella-former with
$N_\text{A}=N_\text{B}/4$ mixed with sphere-formers of various
architectures. The lamella-former fraction $\phi'/\phi$ is
set to
$25\%$. (a) Core radius as a function of the
hydrophilic block length. Points corresponding to simple micelles
(Figure \ref{micelle_fig}) are marked by closed
circles, ABA aggregates (Figure \ref{encapsulated_fig}) or small vesicles
(Figure \ref{vesicle_fig}) are marked with open
circles, and weakly-structured aggregates (Figure \ref{strange_fig}) by asterisks. The data for
the system with a Flory parameter of $\chi N_\text{B}=15$ are connected with
dotted lines; those for the system with $\chi N_\text{B}=22.5$ by
dashed lines. (b) Fraction of the core that is composed of A-blocks for each of the two
systems (upper two curves), and fraction of the core
that is composed of homopolymer solvent (lower two curves). 
}
\end{figure}

The relative insensitivity to sphere-former architecture observed in both
the systems discussed in this section can clearly be seen from plots
of the core radius and composition as a function of the sphere-former
A-block length (Figure \ref{radius_comp_sphere_fig}). The growth of the core
radius shown in Figure 
\ref{radius_comp_sphere_fig}a is clearly weaker than that seen in the
corresponding plots of Figure \ref{radius_comp_conc_fig} and Figure
\ref{radius_comp_lam_fig}. Furthermore, the fraction of the core
composed of A-blocks (upper lines) or solvent (lower lines) remains rather
close to constant, although a weak growth in the amount of solvent in
the open structure of Figure \ref{spherecut45_fig} can be seen. This is in
line with the relatively unchanging morphologies plotted in Figure
\ref{spherecut45_fig} and Figure \ref{spherecut30_fig}.

\section{Conclusions}\label{conclusions}

Using a coarse-grained mean-field approach (self-consistent field theory) we have modeled
several aspects of the formation of small, spherically-symmetric aggregates in a solution of
sphere-forming amphiphile mixed with a smaller amount of
lamella-forming amphiphile. By varying the interaction strength,
architecture and mixing ratio of the amphiphiles, we have found a range of morphologies. When the two
species were similar in architecture, or when only a small admixture
of lamella-forming amphiphile was added, we found simple spherical
micelles with purely hydrophobic cores formed from a mixture of the B-blocks of
the two amphiphiles. For more
strongly mismatched amphiphiles and higher concentrations of
lamella-former, we found complex micelles and small vesicles. Specifically, as the concentration of
lamella-former was gradually increased in a strongly mismatched system with a
relatively high $\chi$ parameter, the simple micelle formed at low
lamella-former concentrations gradually expanded, first forming a more
complex micelle with both A- and B-blocks in the core and then a small
vesicle. For similar systems with lower Flory parameters, the addition of
lamella-former resulted in the formation of a intriguing micellar
structure in which a large and relatively unstructured core of lamella-former is
surrounded by a layer of sphere-forming copolymers. Were this
structure able to be stabilized in experiments, it could prove to be
useful for the solubilization and delivery of hydrophobic compounds, since it
contains a large amount of hydrophobic blocks while retaining a
relatively small size. The formation of these aggregates was shown to
require not only a relatively weak interaction between the two
copolymers, but also for one of the species to have a very short hydrophilic
block. The other complex micelles and small vesicles were present over a much
wider range of lamella-formers. The architecture of the sphere-formers
was found to have a rather weak effect on the aggregate morphology.

The work presented here provides several examples of the wide range of aggregates that may be
formed when two amphiphile species that individually self-assemble
into aggregates of different curvatures are mixed, and gives broad
guidance as to how the polymer parameters might be varied in order to
form a given structure. Furthermore, several of the structures shown
here show the segregation of amphiphiles according to
curvature \cite{sorre,zidovska}. Specifically, in many cases, the sphere-forming amphiphiles
move to the positively-curved surface of the aggregate. Effectively
one-dimensional aggregates such as those considered here are among the
simplest possible systems in which this phenomenon could take place.

Several possible extensions of the current work suggest
themselves. First, given the potential for the solubilization of
hydrophobic chemicals of the large micelles with lamella-former cores,
more realistic interaction parameters and modeling of the polymers (if
necessary by more microscopic simulation methods) could be carried out
in order to search for an experimental parameter range in which these
structures could be formed. Such a study could also investigate
further the
formation of small monodisperse vesicles \cite{zidovska} and bilayers
of preferred curvature \cite{safinya_rev} in binary systems. Second, the
study could be extended to mismatched hydrophobic blocks, to allow
comparison with recent experiments \cite{plestil}. Finally, an analogous investigation could
be performed for the binary triblock copolymer blends \cite{oh,lee} of current
interest in drug delivery applications, where large micelles in mixed
systems are indeed seen \cite{lee}. 

%\bibliography{sphericalrefs}

\begin{thebibliography}{51}
\expandafter\ifx\csname natexlab\endcsname\relax\def\natexlab#1{#1}\fi
\expandafter\ifx\csname bibnamefont\endcsname\relax
  \def\bibnamefont#1{#1}\fi
\expandafter\ifx\csname bibfnamefont\endcsname\relax
  \def\bibfnamefont#1{#1}\fi
\expandafter\ifx\csname citenamefont\endcsname\relax
  \def\citenamefont#1{#1}\fi
\expandafter\ifx\csname url\endcsname\relax
  \def\url#1{\texttt{#1}}\fi
\expandafter\ifx\csname urlprefix\endcsname\relax\def\urlprefix{URL }\fi
\providecommand{\bibinfo}[2]{#2}
\providecommand{\eprint}[2][]{\url{#2}}

\bibitem[{\citenamefont{Jain and Bates}(2003)}]{jain_bates}
\bibinfo{author}{\bibfnamefont{S.}~\bibnamefont{Jain}} \bibnamefont{and}
  \bibinfo{author}{\bibfnamefont{F.~S.} \bibnamefont{Bates}},
  \bibinfo{journal}{Science} \textbf{\bibinfo{volume}{300}},
  \bibinfo{pages}{460} (\bibinfo{year}{2003}).

\bibitem[{\citenamefont{Battaglia and Ryan}(2006)}]{battaglia_ryan}
\bibinfo{author}{\bibfnamefont{G.}~\bibnamefont{Battaglia}} \bibnamefont{and}
  \bibinfo{author}{\bibfnamefont{A.~J.} \bibnamefont{Ryan}},
  \bibinfo{journal}{J. Phys.\ Chem.\ B} \textbf{\bibinfo{volume}{110}},
  \bibinfo{pages}{10272} (\bibinfo{year}{2006}).

\bibitem[{\citenamefont{Smart et~al.}(2010)\citenamefont{Smart, Ryan, Howse,
  and Battaglia}}]{smart}
\bibinfo{author}{\bibfnamefont{T.~P.} \bibnamefont{Smart}},
  \bibinfo{author}{\bibfnamefont{A.~J.} \bibnamefont{Ryan}},
  \bibinfo{author}{\bibfnamefont{J.~R.} \bibnamefont{Howse}}, \bibnamefont{and}
  \bibinfo{author}{\bibfnamefont{G.}~\bibnamefont{Battaglia}},
  \bibinfo{journal}{Langmuir} \textbf{\bibinfo{volume}{26}},
  \bibinfo{pages}{7425} (\bibinfo{year}{2010}).

\bibitem[{\citenamefont{Howse et~al.}(2009)\citenamefont{Howse, Jones,
  Battaglia, Ducker, Leggett, and Ryan}}]{howse}
\bibinfo{author}{\bibfnamefont{J.~R.} \bibnamefont{Howse}},
  \bibinfo{author}{\bibfnamefont{R.~A.~L.} \bibnamefont{Jones}},
  \bibinfo{author}{\bibfnamefont{G.}~\bibnamefont{Battaglia}},
  \bibinfo{author}{\bibfnamefont{R.~E.} \bibnamefont{Ducker}},
  \bibinfo{author}{\bibfnamefont{G.~J.} \bibnamefont{Leggett}},
  \bibnamefont{and} \bibinfo{author}{\bibfnamefont{A.~J.} \bibnamefont{Ryan}},
  \bibinfo{journal}{Nat.\ Mater.} \textbf{\bibinfo{volume}{8}},
  \bibinfo{pages}{507} (\bibinfo{year}{2009}).

\bibitem[{\citenamefont{Kim et~al.}(2005)\citenamefont{Kim, Dalhaimer,
  Christian, and Discher}}]{kim}
\bibinfo{author}{\bibfnamefont{Y.}~\bibnamefont{Kim}},
  \bibinfo{author}{\bibfnamefont{P.}~\bibnamefont{Dalhaimer}},
  \bibinfo{author}{\bibfnamefont{D.~A.} \bibnamefont{Christian}},
  \bibnamefont{and} \bibinfo{author}{\bibfnamefont{D.~E.}
  \bibnamefont{Discher}}, \bibinfo{journal}{Nanotechnology}
  \textbf{\bibinfo{volume}{16}}, \bibinfo{pages}{S484} (\bibinfo{year}{2005}).

\bibitem[{\citenamefont{Lomas et~al.}(2007)\citenamefont{Lomas, Canton,
  MacNeil, Du, Armes, Ryan, Lewis, and Battaglia}}]{lomas}
\bibinfo{author}{\bibfnamefont{H.}~\bibnamefont{Lomas}},
  \bibinfo{author}{\bibfnamefont{I.}~\bibnamefont{Canton}},
  \bibinfo{author}{\bibfnamefont{S.}~\bibnamefont{MacNeil}},
  \bibinfo{author}{\bibfnamefont{J.}~\bibnamefont{Du}},
  \bibinfo{author}{\bibfnamefont{S.~P.} \bibnamefont{Armes}},
  \bibinfo{author}{\bibfnamefont{A.~J.} \bibnamefont{Ryan}},
  \bibinfo{author}{\bibfnamefont{A.~L.} \bibnamefont{Lewis}}, \bibnamefont{and}
  \bibinfo{author}{\bibfnamefont{G.}~\bibnamefont{Battaglia}},
  \bibinfo{journal}{Adv.\ Mater.} \textbf{\bibinfo{volume}{19}},
  \bibinfo{pages}{4238} (\bibinfo{year}{2007}).

\bibitem[{\citenamefont{Zidovska et~al.}(2009)\citenamefont{Zidovska, Ewert,
  Quispe, Carragher, Potter, and Safinya}}]{zidovska}
\bibinfo{author}{\bibfnamefont{A.}~\bibnamefont{Zidovska}},
  \bibinfo{author}{\bibfnamefont{K.~K.} \bibnamefont{Ewert}},
  \bibinfo{author}{\bibfnamefont{J.}~\bibnamefont{Quispe}},
  \bibinfo{author}{\bibfnamefont{B.}~\bibnamefont{Carragher}},
  \bibinfo{author}{\bibfnamefont{C.~S.} \bibnamefont{Potter}},
  \bibnamefont{and} \bibinfo{author}{\bibfnamefont{C.~R.}
  \bibnamefont{Safinya}}, \bibinfo{journal}{Langmuir}
  \textbf{\bibinfo{volume}{25}}, \bibinfo{pages}{2979} (\bibinfo{year}{2009}).

\bibitem[{\citenamefont{Kinning et~al.}(1988)\citenamefont{Kinning, Winey, and
  Thomas}}]{kinning_winey_thomas}
\bibinfo{author}{\bibfnamefont{D.~J.} \bibnamefont{Kinning}},
  \bibinfo{author}{\bibfnamefont{K.~I.} \bibnamefont{Winey}}, \bibnamefont{and}
  \bibinfo{author}{\bibfnamefont{E.~L.} \bibnamefont{Thomas}},
  \bibinfo{journal}{Macromolecules} \textbf{\bibinfo{volume}{21}},
  \bibinfo{pages}{3502} (\bibinfo{year}{1988}).

\bibitem[{\citenamefont{Israelachvili et~al.}(1976)\citenamefont{Israelachvili,
  Mitchell, and Ninham}}]{israelachvili}
\bibinfo{author}{\bibfnamefont{J.~N.} \bibnamefont{Israelachvili}},
  \bibinfo{author}{\bibfnamefont{D.~J.} \bibnamefont{Mitchell}},
  \bibnamefont{and} \bibinfo{author}{\bibfnamefont{B.~W.}
  \bibnamefont{Ninham}}, \bibinfo{journal}{J. Chem.\ Soc.\ Faraday Trans.\ 1}
  \textbf{\bibinfo{volume}{72}}, \bibinfo{pages}{1525} (\bibinfo{year}{1976}).

\bibitem[{\citenamefont{Adams et~al.}(2009)\citenamefont{Adams, Kitchen, Adams,
  Furzeland, Atkins, Schuetz, Fernyhough, Tzokova, Ryan, and Butler}}]{adams}
\bibinfo{author}{\bibfnamefont{D.~J.} \bibnamefont{Adams}},
  \bibinfo{author}{\bibfnamefont{C.}~\bibnamefont{Kitchen}},
  \bibinfo{author}{\bibfnamefont{S.}~\bibnamefont{Adams}},
  \bibinfo{author}{\bibfnamefont{S.}~\bibnamefont{Furzeland}},
  \bibinfo{author}{\bibfnamefont{D.}~\bibnamefont{Atkins}},
  \bibinfo{author}{\bibfnamefont{P.}~\bibnamefont{Schuetz}},
  \bibinfo{author}{\bibfnamefont{C.~M.} \bibnamefont{Fernyhough}},
  \bibinfo{author}{\bibfnamefont{N.}~\bibnamefont{Tzokova}},
  \bibinfo{author}{\bibfnamefont{A.~J.} \bibnamefont{Ryan}}, \bibnamefont{and}
  \bibinfo{author}{\bibfnamefont{M.~F.} \bibnamefont{Butler}},
  \bibinfo{journal}{Soft Matter} \textbf{\bibinfo{volume}{5}},
  \bibinfo{pages}{3086} (\bibinfo{year}{2009}).

\bibitem[{\citenamefont{Kaya et~al.}(2002)\citenamefont{Kaya, Willner,
  Allgaier, and Richter}}]{kaya}
\bibinfo{author}{\bibfnamefont{H.}~\bibnamefont{Kaya}},
  \bibinfo{author}{\bibfnamefont{L.}~\bibnamefont{Willner}},
  \bibinfo{author}{\bibfnamefont{J.}~\bibnamefont{Allgaier}}, \bibnamefont{and}
  \bibinfo{author}{\bibfnamefont{D.}~\bibnamefont{Richter}},
  \bibinfo{journal}{App.\ Phys.\ A: Mater.\ Sci.\ Process.}
  \textbf{\bibinfo{volume}{74}}, \bibinfo{pages}{S499} (\bibinfo{year}{2002}).

\bibitem[{\citenamefont{Schuetz et~al.}(2011)\citenamefont{Schuetz, Greenall,
  Bent, Furzeland, Atkins, Butler, McLeish, and Buzza}}]{schuetz}
\bibinfo{author}{\bibfnamefont{P.}~\bibnamefont{Schuetz}},
  \bibinfo{author}{\bibfnamefont{M.~J.} \bibnamefont{Greenall}},
  \bibinfo{author}{\bibfnamefont{J.}~\bibnamefont{Bent}},
  \bibinfo{author}{\bibfnamefont{S.}~\bibnamefont{Furzeland}},
  \bibinfo{author}{\bibfnamefont{D.}~\bibnamefont{Atkins}},
  \bibinfo{author}{\bibfnamefont{M.~F.} \bibnamefont{Butler}},
  \bibinfo{author}{\bibfnamefont{T.~C.~B.} \bibnamefont{McLeish}},
  \bibnamefont{and} \bibinfo{author}{\bibfnamefont{D.~M.~A.}
  \bibnamefont{Buzza}}, \bibinfo{journal}{Soft Matter}
  \textbf{\bibinfo{volume}{7}}, \bibinfo{pages}{749} (\bibinfo{year}{2011}).

\bibitem[{\citenamefont{Dan and Safran}(2006)}]{dan_safran}
\bibinfo{author}{\bibfnamefont{N.}~\bibnamefont{Dan}} \bibnamefont{and}
  \bibinfo{author}{\bibfnamefont{S.~A.} \bibnamefont{Safran}},
  \bibinfo{journal}{Adv. Colloid Interface Sci.}
  \textbf{\bibinfo{volume}{123}}, \bibinfo{pages}{323} (\bibinfo{year}{2006}).

\bibitem[{\citenamefont{Kaler et~al.}(1989)\citenamefont{Kaler, Murthy,
  Rodriguez, and Zasadzinski}}]{kaler}
\bibinfo{author}{\bibfnamefont{E.~W.} \bibnamefont{Kaler}},
  \bibinfo{author}{\bibfnamefont{A.~K.} \bibnamefont{Murthy}},
  \bibinfo{author}{\bibfnamefont{B.~E.} \bibnamefont{Rodriguez}},
  \bibnamefont{and} \bibinfo{author}{\bibfnamefont{J.~A.~N.}
  \bibnamefont{Zasadzinski}}, \bibinfo{journal}{Science}
  \textbf{\bibinfo{volume}{245}}, \bibinfo{pages}{1371} (\bibinfo{year}{1989}).

\bibitem[{\citenamefont{Jain and Bates}(2004)}]{jain_bates_macro}
\bibinfo{author}{\bibfnamefont{S.}~\bibnamefont{Jain}} \bibnamefont{and}
  \bibinfo{author}{\bibfnamefont{F.~S.} \bibnamefont{Bates}},
  \bibinfo{journal}{Macromolecules} \textbf{\bibinfo{volume}{37}},
  \bibinfo{pages}{1511} (\bibinfo{year}{2004}).

\bibitem[{\citenamefont{Safran et~al.}(1990)\citenamefont{Safran, Pincus, and
  Andelman}}]{safran_pincus_andelman}
\bibinfo{author}{\bibfnamefont{S.~A.} \bibnamefont{Safran}},
  \bibinfo{author}{\bibfnamefont{P.}~\bibnamefont{Pincus}}, \bibnamefont{and}
  \bibinfo{author}{\bibfnamefont{D.}~\bibnamefont{Andelman}},
  \bibinfo{journal}{Science} \textbf{\bibinfo{volume}{248}},
  \bibinfo{pages}{354} (\bibinfo{year}{1990}).

\bibitem[{\citenamefont{Safran et~al.}(1991)\citenamefont{Safran, Pincus,
  Andelman, and MacKintosh}}]{safran_et_al}
\bibinfo{author}{\bibfnamefont{S.~A.} \bibnamefont{Safran}},
  \bibinfo{author}{\bibfnamefont{P.~A.} \bibnamefont{Pincus}},
  \bibinfo{author}{\bibfnamefont{D.}~\bibnamefont{Andelman}}, \bibnamefont{and}
  \bibinfo{author}{\bibfnamefont{F.~C.} \bibnamefont{MacKintosh}},
  \bibinfo{journal}{Phys.\ Rev.\ A} \textbf{\bibinfo{volume}{43}},
  \bibinfo{pages}{1071} (\bibinfo{year}{1991}).

\bibitem[{\citenamefont{Vinson et~al.}(1989)\citenamefont{Vinson, Talmon, and
  Walter}}]{vinson}
\bibinfo{author}{\bibfnamefont{P.~K.} \bibnamefont{Vinson}},
  \bibinfo{author}{\bibfnamefont{Y.}~\bibnamefont{Talmon}}, \bibnamefont{and}
  \bibinfo{author}{\bibfnamefont{A.}~\bibnamefont{Walter}},
  \bibinfo{journal}{Biophys.\ J.} \textbf{\bibinfo{volume}{56}},
  \bibinfo{pages}{669} (\bibinfo{year}{1989}).

\bibitem[{\citenamefont{Oberdisse et~al.}(1998)\citenamefont{Oberdisse, Regev,
  and Porte}}]{oberdisse}
\bibinfo{author}{\bibfnamefont{J.}~\bibnamefont{Oberdisse}},
  \bibinfo{author}{\bibfnamefont{O.}~\bibnamefont{Regev}}, \bibnamefont{and}
  \bibinfo{author}{\bibfnamefont{G.}~\bibnamefont{Porte}}, \bibinfo{journal}{J.
  Phys.\ Chem.\ B} \textbf{\bibinfo{volume}{102}}, \bibinfo{pages}{1102}
  (\bibinfo{year}{1998}).

\bibitem[{\citenamefont{Sorre et~al.}(2009)\citenamefont{Sorre, Callan-Jones,
  Manneville, Nassoy, Joanny, Prost, Goud, and Bassereau}}]{sorre}
\bibinfo{author}{\bibfnamefont{B.}~\bibnamefont{Sorre}},
  \bibinfo{author}{\bibfnamefont{A.}~\bibnamefont{Callan-Jones}},
  \bibinfo{author}{\bibfnamefont{J.~B.} \bibnamefont{Manneville}},
  \bibinfo{author}{\bibfnamefont{P.}~\bibnamefont{Nassoy}},
  \bibinfo{author}{\bibfnamefont{J.~F.} \bibnamefont{Joanny}},
  \bibinfo{author}{\bibfnamefont{J.}~\bibnamefont{Prost}},
  \bibinfo{author}{\bibfnamefont{B.}~\bibnamefont{Goud}}, \bibnamefont{and}
  \bibinfo{author}{\bibfnamefont{P.}~\bibnamefont{Bassereau}},
  \bibinfo{journal}{Proc.\ Natl.\ Acad.\ Sci.\ U.S.A}
  \textbf{\bibinfo{volume}{106}}, \bibinfo{pages}{5622} (\bibinfo{year}{2009}).

\bibitem[{\citenamefont{Greenall and Gompper}(2011)}]{gg}
\bibinfo{author}{\bibfnamefont{M.~J.} \bibnamefont{Greenall}} \bibnamefont{and}
  \bibinfo{author}{\bibfnamefont{G.}~\bibnamefont{Gompper}},
  \bibinfo{journal}{Langmuir} \textbf{\bibinfo{volume}{27}},
  \bibinfo{pages}{3416} (\bibinfo{year}{2011}).

\bibitem[{\citenamefont{Akiyoshi et~al.}(2003)\citenamefont{Akiyoshi, Itaya,
  Nomura, Ono, and Yoshikawa}}]{akiyoshi}
\bibinfo{author}{\bibfnamefont{K.}~\bibnamefont{Akiyoshi}},
  \bibinfo{author}{\bibfnamefont{A.}~\bibnamefont{Itaya}},
  \bibinfo{author}{\bibfnamefont{S.~M.} \bibnamefont{Nomura}},
  \bibinfo{author}{\bibfnamefont{N.}~\bibnamefont{Ono}}, \bibnamefont{and}
  \bibinfo{author}{\bibfnamefont{K.}~\bibnamefont{Yoshikawa}},
  \bibinfo{journal}{FEBS Lett.} \textbf{\bibinfo{volume}{534}},
  \bibinfo{pages}{33} (\bibinfo{year}{2003}).

\bibitem[{\citenamefont{Lee et~al.}(2011)\citenamefont{Lee, Oh, Youn, Nam,
  Park, Yun, Kim, Song, and Oh}}]{lee}
\bibinfo{author}{\bibfnamefont{E.~S.} \bibnamefont{Lee}},
  \bibinfo{author}{\bibfnamefont{Y.~T.} \bibnamefont{Oh}},
  \bibinfo{author}{\bibfnamefont{Y.~S.} \bibnamefont{Youn}},
  \bibinfo{author}{\bibfnamefont{M.}~\bibnamefont{Nam}},
  \bibinfo{author}{\bibfnamefont{B.}~\bibnamefont{Park}},
  \bibinfo{author}{\bibfnamefont{J.}~\bibnamefont{Yun}},
  \bibinfo{author}{\bibfnamefont{J.~H.} \bibnamefont{Kim}},
  \bibinfo{author}{\bibfnamefont{H.-T.} \bibnamefont{Song}}, \bibnamefont{and}
  \bibinfo{author}{\bibfnamefont{K.~T.} \bibnamefont{Oh}},
  \bibinfo{journal}{Colloids Surf. B} \textbf{\bibinfo{volume}{82}},
  \bibinfo{pages}{190} (\bibinfo{year}{2011}).

\bibitem[{\citenamefont{Oh et~al.}(2004)\citenamefont{Oh, Bronich, and
  Kabanov}}]{oh}
\bibinfo{author}{\bibfnamefont{K.~T.} \bibnamefont{Oh}},
  \bibinfo{author}{\bibfnamefont{T.~K.} \bibnamefont{Bronich}},
  \bibnamefont{and} \bibinfo{author}{\bibfnamefont{A.~V.}
  \bibnamefont{Kabanov}}, \bibinfo{journal}{J. Controlled Release}
  \textbf{\bibinfo{volume}{94}}, \bibinfo{pages}{411} (\bibinfo{year}{2004}).

\bibitem[{\citenamefont{Edwards}(1965)}]{edwards}
\bibinfo{author}{\bibfnamefont{S.~F.} \bibnamefont{Edwards}},
  \bibinfo{journal}{Proc.\ Phys.\ Soc.} \textbf{\bibinfo{volume}{85}},
  \bibinfo{pages}{613} (\bibinfo{year}{1965}).

\bibitem[{\citenamefont{Maniadis et~al.}(2007)\citenamefont{Maniadis, Lookman,
  Kober, and Rasmussen}}]{maniadis}
\bibinfo{author}{\bibfnamefont{P.}~\bibnamefont{Maniadis}},
  \bibinfo{author}{\bibfnamefont{T.}~\bibnamefont{Lookman}},
  \bibinfo{author}{\bibfnamefont{E.~M.} \bibnamefont{Kober}}, \bibnamefont{and}
  \bibinfo{author}{\bibfnamefont{K.~O.} \bibnamefont{Rasmussen}},
  \bibinfo{journal}{Phys.\ Rev.\ Lett.} \textbf{\bibinfo{volume}{99}},
  \bibinfo{pages}{048302} (\bibinfo{year}{2007}).

\bibitem[{\citenamefont{Drolet and Fredrickson}(1999)}]{drolet_fredrickson}
\bibinfo{author}{\bibfnamefont{F.}~\bibnamefont{Drolet}} \bibnamefont{and}
  \bibinfo{author}{\bibfnamefont{G.~H.} \bibnamefont{Fredrickson}},
  \bibinfo{journal}{Phys.\ Rev.\ Lett.} \textbf{\bibinfo{volume}{83}},
  \bibinfo{pages}{4317} (\bibinfo{year}{1999}).

\bibitem[{\citenamefont{Matsen}(2006)}]{matsen_book}
\bibinfo{author}{\bibfnamefont{M.~W.} \bibnamefont{Matsen}},
  \emph{\bibinfo{title}{Soft Matter}} (\bibinfo{publisher}{Wiley-VCH},
  \bibinfo{address}{Weinheim}, \bibinfo{year}{2006}),
  chap.~\bibinfo{chapter}{2}.

\bibitem[{\citenamefont{Duque}(2003)}]{duque}
\bibinfo{author}{\bibfnamefont{D.}~\bibnamefont{Duque}}, \bibinfo{journal}{J.
  Chem.\ Phys.} \textbf{\bibinfo{volume}{119}}, \bibinfo{pages}{5701}
  (\bibinfo{year}{2003}).

\bibitem[{\citenamefont{Katsov et~al.}(2004)\citenamefont{Katsov, M\"{u}ller,
  and Schick}}]{katsov1}
\bibinfo{author}{\bibfnamefont{K.}~\bibnamefont{Katsov}},
  \bibinfo{author}{\bibfnamefont{M.}~\bibnamefont{M\"{u}ller}},
  \bibnamefont{and} \bibinfo{author}{\bibfnamefont{M.}~\bibnamefont{Schick}},
  \bibinfo{journal}{Biophys.\ J.} \textbf{\bibinfo{volume}{87}},
  \bibinfo{pages}{3277} (\bibinfo{year}{2004}).

\bibitem[{\citenamefont{Cavallo et~al.}(2006)\citenamefont{Cavallo, M\"{u}ller,
  and Binder}}]{cavallo}
\bibinfo{author}{\bibfnamefont{A.}~\bibnamefont{Cavallo}},
  \bibinfo{author}{\bibfnamefont{M.}~\bibnamefont{M\"{u}ller}},
  \bibnamefont{and} \bibinfo{author}{\bibfnamefont{K.}~\bibnamefont{Binder}},
  \bibinfo{journal}{Macromolecules} \textbf{\bibinfo{volume}{39}},
  \bibinfo{pages}{9539} (\bibinfo{year}{2006}).

\bibitem[{\citenamefont{Wijmans and Linse}(1995)}]{wijmans_linse}
\bibinfo{author}{\bibfnamefont{C.~M.} \bibnamefont{Wijmans}} \bibnamefont{and}
  \bibinfo{author}{\bibfnamefont{P.}~\bibnamefont{Linse}},
  \bibinfo{journal}{Langmuir} \textbf{\bibinfo{volume}{11}},
  \bibinfo{pages}{3748} (\bibinfo{year}{1995}).

\bibitem[{\citenamefont{Leermakers and
  Scheutjens}(1990)}]{leermakers_scheutjens-shape}
\bibinfo{author}{\bibfnamefont{F.~A.~M.} \bibnamefont{Leermakers}}
  \bibnamefont{and} \bibinfo{author}{\bibfnamefont{J.~M. H.~M.}
  \bibnamefont{Scheutjens}}, \bibinfo{journal}{J. Colloid Interface Sci.}
  \textbf{\bibinfo{volume}{136}}, \bibinfo{pages}{231} (\bibinfo{year}{1990}).

\bibitem[{\citenamefont{Schmid}(1998)}]{schmid_scf_rev}
\bibinfo{author}{\bibfnamefont{F.}~\bibnamefont{Schmid}}, \bibinfo{journal}{J.
  Phys.: Condens.\ Matter} \textbf{\bibinfo{volume}{10}}, \bibinfo{pages}{8105}
  (\bibinfo{year}{1998}).

\bibitem[{\citenamefont{Jones}(2002)}]{jones_book}
\bibinfo{author}{\bibfnamefont{R.~A.~L.} \bibnamefont{Jones}},
  \emph{\bibinfo{title}{Soft Condensed Matter}} (\bibinfo{publisher}{Oxford
  University Press}, \bibinfo{address}{Oxford}, \bibinfo{year}{2002}).

\bibitem[{\citenamefont{Werner et~al.}(1999)\citenamefont{Werner, M\"uller,
  Schmid, and Binder}}]{werner}
\bibinfo{author}{\bibfnamefont{A.}~\bibnamefont{Werner}},
  \bibinfo{author}{\bibfnamefont{M.}~\bibnamefont{M\"uller}},
  \bibinfo{author}{\bibfnamefont{F.}~\bibnamefont{Schmid}}, \bibnamefont{and}
  \bibinfo{author}{\bibfnamefont{K.}~\bibnamefont{Binder}},
  \bibinfo{journal}{J. Chem.\ Phys.} \textbf{\bibinfo{volume}{110}},
  \bibinfo{pages}{1221} (\bibinfo{year}{1999}).

\bibitem[{\citenamefont{M\"{u}ller and Gompper}(2002)}]{mueller}
\bibinfo{author}{\bibfnamefont{M.}~\bibnamefont{M\"{u}ller}} \bibnamefont{and}
  \bibinfo{author}{\bibfnamefont{G.}~\bibnamefont{Gompper}},
  \bibinfo{journal}{Phys.\ Rev.\ E} \textbf{\bibinfo{volume}{66}},
  \bibinfo{pages}{041805} (\bibinfo{year}{2002}).

\bibitem[{\citenamefont{Wang et~al.}(2010)\citenamefont{Wang, Guo, An,
  M\"uller, and Wang}}]{wang}
\bibinfo{author}{\bibfnamefont{J.~F.} \bibnamefont{Wang}},
  \bibinfo{author}{\bibfnamefont{K.~K.} \bibnamefont{Guo}},
  \bibinfo{author}{\bibfnamefont{L.~J.} \bibnamefont{An}},
  \bibinfo{author}{\bibfnamefont{M.}~\bibnamefont{M\"uller}}, \bibnamefont{and}
  \bibinfo{author}{\bibfnamefont{Z.~G.} \bibnamefont{Wang}},
  \bibinfo{journal}{Macromolecules} \textbf{\bibinfo{volume}{43}},
  \bibinfo{pages}{2037} (\bibinfo{year}{2010}).

\bibitem[{\citenamefont{Denesyuk and Gompper}(2006)}]{denesyuk}
\bibinfo{author}{\bibfnamefont{N.~A.} \bibnamefont{Denesyuk}} \bibnamefont{and}
  \bibinfo{author}{\bibfnamefont{G.}~\bibnamefont{Gompper}},
  \bibinfo{journal}{Macromolecules} \textbf{\bibinfo{volume}{39}},
  \bibinfo{pages}{5497} (\bibinfo{year}{2006}).

\bibitem[{\citenamefont{Mayes and Delacruz}(1988)}]{mayes}
\bibinfo{author}{\bibfnamefont{A.~M.} \bibnamefont{Mayes}} \bibnamefont{and}
  \bibinfo{author}{\bibfnamefont{M.~O.} \bibnamefont{Delacruz}},
  \bibinfo{journal}{Macromolecules} \textbf{\bibinfo{volume}{21}},
  \bibinfo{pages}{2543} (\bibinfo{year}{1988}).

\bibitem[{\citenamefont{Fredrickson}(2006)}]{fredrickson_book}
\bibinfo{author}{\bibfnamefont{G.~H.} \bibnamefont{Fredrickson}},
  \emph{\bibinfo{title}{The Equilibrium Theory of Inhomogeneous Polymers}}
  (\bibinfo{publisher}{Oxford University Press}, \bibinfo{address}{Oxford},
  \bibinfo{year}{2006}).

\bibitem[{\citenamefont{Matsen}(2004)}]{matsen2004}
\bibinfo{author}{\bibfnamefont{M.~W.} \bibnamefont{Matsen}},
  \bibinfo{journal}{J. Chem.\ Phys.} \textbf{\bibinfo{volume}{121}},
  \bibinfo{pages}{1938} (\bibinfo{year}{2004}).

\bibitem[{\citenamefont{Press et~al.}(1992)\citenamefont{Press, Flannery,
  Teukolsky, and Vetterling}}]{num_rec}
\bibinfo{author}{\bibfnamefont{W.~H.} \bibnamefont{Press}},
  \bibinfo{author}{\bibfnamefont{B.~P.} \bibnamefont{Flannery}},
  \bibinfo{author}{\bibfnamefont{S.~A.} \bibnamefont{Teukolsky}},
  \bibnamefont{and} \bibinfo{author}{\bibfnamefont{W.~T.}
  \bibnamefont{Vetterling}}, \emph{\bibinfo{title}{Numerical Recipes in C}}
  (\bibinfo{publisher}{Cambridge University Press},
  \bibinfo{address}{Cambridge}, \bibinfo{year}{1992}), \bibinfo{edition}{2nd}
  ed.

\bibitem[{\citenamefont{Greenall
  et~al.}(2009{\natexlab{a}})\citenamefont{Greenall, Buzza, and
  McLeish}}]{gbm_macro}
\bibinfo{author}{\bibfnamefont{M.~J.} \bibnamefont{Greenall}},
  \bibinfo{author}{\bibfnamefont{D.~M.~A.} \bibnamefont{Buzza}},
  \bibnamefont{and} \bibinfo{author}{\bibfnamefont{T.~C.~B.}
  \bibnamefont{McLeish}}, \bibinfo{journal}{Macromolecules}
  \textbf{\bibinfo{volume}{42}}, \bibinfo{pages}{5873}
  (\bibinfo{year}{2009}{\natexlab{a}}).

\bibitem[{\citenamefont{Greenall
  et~al.}(2009{\natexlab{b}})\citenamefont{Greenall, Buzza, and
  McLeish}}]{gbm_jcp}
\bibinfo{author}{\bibfnamefont{M.~J.} \bibnamefont{Greenall}},
  \bibinfo{author}{\bibfnamefont{D.~M.~A.} \bibnamefont{Buzza}},
  \bibnamefont{and} \bibinfo{author}{\bibfnamefont{T.~C.~B.}
  \bibnamefont{McLeish}}, \bibinfo{journal}{J. Chem.\ Phys.}
  \textbf{\bibinfo{volume}{131}}, \bibinfo{pages}{034904}
  (\bibinfo{year}{2009}{\natexlab{b}}).

\bibitem[{\citenamefont{Shim et~al.}(1991)\citenamefont{Shim, Marques, and
  Cates}}]{shim}
\bibinfo{author}{\bibfnamefont{D.~F.~K.} \bibnamefont{Shim}},
  \bibinfo{author}{\bibfnamefont{C.}~\bibnamefont{Marques}}, \bibnamefont{and}
  \bibinfo{author}{\bibfnamefont{M.~E.} \bibnamefont{Cates}},
  \bibinfo{journal}{Macromolecules} \textbf{\bibinfo{volume}{24}},
  \bibinfo{pages}{5309} (\bibinfo{year}{1991}).

\bibitem[{\citenamefont{Li et~al.}(2009{\natexlab{a}})\citenamefont{Li,
  Marcelis, Sudholter, Stuart, and Leermakers}}]{li}
\bibinfo{author}{\bibfnamefont{F.}~\bibnamefont{Li}},
  \bibinfo{author}{\bibfnamefont{A.~T.~M.} \bibnamefont{Marcelis}},
  \bibinfo{author}{\bibfnamefont{E.~J.~R.} \bibnamefont{Sudholter}},
  \bibinfo{author}{\bibfnamefont{M.~A.~C.} \bibnamefont{Stuart}},
  \bibnamefont{and} \bibinfo{author}{\bibfnamefont{F.~A.~M.}
  \bibnamefont{Leermakers}}, \bibinfo{journal}{Soft Matter}
  \textbf{\bibinfo{volume}{5}}, \bibinfo{pages}{4173}
  (\bibinfo{year}{2009}{\natexlab{a}}).

\bibitem[{\citenamefont{Li et~al.}(2009{\natexlab{b}})\citenamefont{Li,
  Prevost, Schweins, Marcelis, Leermakers, Stuart, and Sudholter}}]{li2}
\bibinfo{author}{\bibfnamefont{F.}~\bibnamefont{Li}},
  \bibinfo{author}{\bibfnamefont{S.}~\bibnamefont{Prevost}},
  \bibinfo{author}{\bibfnamefont{R.}~\bibnamefont{Schweins}},
  \bibinfo{author}{\bibfnamefont{A.~T.~M.} \bibnamefont{Marcelis}},
  \bibinfo{author}{\bibfnamefont{F.~A.~M.} \bibnamefont{Leermakers}},
  \bibinfo{author}{\bibfnamefont{M.~A.~C.} \bibnamefont{Stuart}},
  \bibnamefont{and} \bibinfo{author}{\bibfnamefont{E.~J.~R.}
  \bibnamefont{Sudholter}}, \bibinfo{journal}{Soft Matter}
  \textbf{\bibinfo{volume}{5}}, \bibinfo{pages}{4169}
  (\bibinfo{year}{2009}{\natexlab{b}}).

\bibitem[{\citenamefont{Safinya}(1997)}]{safinya_rev}
\bibinfo{author}{\bibfnamefont{C.~R.} \bibnamefont{Safinya}},
  \bibinfo{journal}{Colloids Surf. A} \textbf{\bibinfo{volume}{128}},
  \bibinfo{pages}{183} (\bibinfo{year}{1997}).

\bibitem[{\citenamefont{Cooke and Deserno}(2006)}]{cooke_deserno}
\bibinfo{author}{\bibfnamefont{I.~R.} \bibnamefont{Cooke}} \bibnamefont{and}
  \bibinfo{author}{\bibfnamefont{M.}~\bibnamefont{Deserno}},
  \bibinfo{journal}{Biophys.\ J.} \textbf{\bibinfo{volume}{91}},
  \bibinfo{pages}{487} (\bibinfo{year}{2006}).

\bibitem[{\citenamefont{Ple\v{s}til et~al.}(2006)\citenamefont{Ple\v{s}til,
  Ko\v{n}\'{a}k, Ju, and Lal}}]{plestil}
\bibinfo{author}{\bibfnamefont{J.}~\bibnamefont{Ple\v{s}til}},
  \bibinfo{author}{\bibfnamefont{C.}~\bibnamefont{Ko\v{n}\'{a}k}},
  \bibinfo{author}{\bibfnamefont{X.}~\bibnamefont{Ju}}, \bibnamefont{and}
  \bibinfo{author}{\bibfnamefont{J.}~\bibnamefont{Lal}},
  \bibinfo{journal}{Macromol.\ Chem.\ Phys.} \textbf{\bibinfo{volume}{207}},
  \bibinfo{pages}{231} (\bibinfo{year}{2006}).

\end{thebibliography}

\end{document}